\documentclass[twocolumn,superscriptaddress,aps,prl,showpacs,preprintnumbers,amsmath,amssymb,floatfix]{revtex4-1}

\usepackage[latin9]{inputenc}
\usepackage{color}
\usepackage{dcolumn}
\usepackage{bm}
\usepackage{bbm}
\usepackage{braket}
\usepackage{mathrsfs}
\usepackage{amstext}

\usepackage{libertine}
\usepackage{txfonts}

\usepackage{amsmath}
\usepackage{dsfont}
\usepackage{euscript}
\usepackage{amssymb}
\usepackage{graphicx}
\usepackage{esint}

\usepackage[unicode=true,
 bookmarks=false,
 breaklinks=false,pdfborder={0 0 1},backref=false,colorlinks=true,citecolor=blue]
 {hyperref}

\makeatletter
\@ifundefined{textcolor}{}
{%
 \definecolor{BLACK}{gray}{0}
 \definecolor{WHITE}{gray}{1}
 \definecolor{RED}{rgb}{1,0,0}
 \definecolor{GREEN}{rgb}{0,1,0}
 \definecolor{BLUE}{rgb}{0,0,1}
 \definecolor{CYAN}{cmyk}{1,0,0,0}
 \definecolor{MAGENTA}{cmyk}{0,1,0,0}
 \definecolor{YELLOW}{cmyk}{0,0,1,0}
}

\newcommand{\eref}[1]{Eq.\,\eqref{#1}}
\newcommand{\fref}[1]{Fig.\,\ref{#1}}
\newcommand{\sref}[1]{Sec.\,\ref{#1}}
\newcommand{\aref}[1]{Appendix \ref{#1}}

\newcommand{\mc}[1]{\mathcal{#1}}

\makeatother

\begin{document}

\title{Emergent equilibrium in many-body optical bistability}

\author{M.\,Foss-Feig}
\affiliation{United States Army Research Laboratory, Adelphi, MD 20783, USA}
\affiliation{Joint Quantum Institute, NIST/University of Maryland, College Park, MD 20742 USA}
\affiliation{Joint Center for Quantum Information and Computer Science, NIST/University of Maryland, College Park, MD 20742 USA}

\author{P.\,Niroula}
\affiliation{Joint Quantum Institute, NIST/University of Maryland, College Park, MD 20742 USA}
\affiliation{Department of Physics, Harvard University, Cambridge, MA 02138, USA}

\author{J.\,T.\,Young}
\affiliation{Joint Quantum Institute, NIST/University of Maryland, College Park, MD 20742 USA}

\author{M.\,Hafezi}
\affiliation{Joint Quantum Institute, NIST/University of Maryland, College Park, MD 20742 USA}
\affiliation{Department of Electrical and Computer Engineering and Institute for Research in Electronics and Applied Physics, University of Maryland, College Park, MD 20742, USA}

\author{A.\,V.\,Gorshkov}
\affiliation{Joint Quantum Institute, NIST/University of Maryland, College Park, MD 20742 USA}
\affiliation{Joint Center for Quantum Information and Computer Science, NIST/University of Maryland, College Park, MD 20742 USA}

\author{R.\,M.\,Wilson}
\affiliation{Department of Physics, The United States Naval Academy, Annapolis, MD 21402, USA}

\author{M.\,F.\,Maghrebi}
\affiliation{Joint Quantum Institute, NIST/University of Maryland, College Park, MD 20742 USA}
\affiliation{Joint Center for Quantum Information and Computer Science, NIST/University of Maryland, College Park, MD 20742 USA}

\begin{abstract}

Many-body systems constructed of quantum-optical building blocks can now be realized in experimental platforms ranging from exciton-polariton fluids to ultracold gases of Rydberg atoms, establishing a fascinating interface between traditional many-body physics and the driven-dissipative, non-equilibrium setting of cavity-QED.  At this interface, the standard techniques and intuitions of both fields are called into question, obscuring issues as fundamental as the role of fluctuations, dimensionality, and symmetry on the nature of collective behavior and phase transitions.  Here, we study the driven-dissipative Bose-Hubbard model, a minimal description of numerous atomic, optical, and solid-state systems in which particle loss is countered by coherent driving.  Despite being a lattice version of optical bistability---a foundational and patently \emph{non-equilibrium} model of cavity-QED---the steady state possesses an emergent \emph{equilibrium} description in terms of a classical Ising model.  We establish this picture by identifying a limit in which the quantum dynamics is asymptotically equivalent to non-equilibrium Langevin equations, which support a phase transition described by model A of the Hohenberg-Halperin classification.  Numerical simulations of the Langevin equations corroborate this picture, producing results consistent with the behavior of a finite-temperature Ising model.

\end{abstract}







\maketitle

\section{Introduction \label{sec: intro}}

While cavity-QED systems often contain many interacting degrees of freedom, they are unconventional from the standpoint of traditional many-body physics for two primary reasons.  First, the mediation of interactions through a small number of delocalized cavity modes generally leads to extremely long-ranged interactions \cite{PhysRev.93.99}, which suppress the role of fluctuations and often enable accurate mean-field descriptions.  In this sense they are simpler than conventional solid-state realizations of many-body physics, in which short-range interactions promote both quantum and thermal fluctuations to an important role, especially in low spatial dimensions \cite{book_auerbach,book_giamarchi}.  Second, cavity-QED systems are typically driven and dissipative; as a result, even if they reach a time-independent steady state they will generally not be in thermal equilibrium  \cite{book_carmichael_1}.  In this sense they are more complicated than conventional solid-state realizations of many-body physics, in which coupling to a thermal reservoir is typically assumed and well justified, leaving the system in thermal equilibrium and enabling the powerful tools of statistical mechanics to be employed \cite{KardarBook}.

In recent years, experimental advances in quantum optics have begun to blur the first of these distinctions \cite{RevModPhys.85.299,0034.4885.79.9.096001,Angelakis_review_2016,Hartmann_review_2016}, with platforms including exciton-polariton fluids in semiconductor quantum wells \cite{Deng199,Kasprazak2006,PhysRevLett.96.230602,Byrnes2014}, circuit-QED \cite{10.1038/nphys2251,PhysRevLett.108.233603,PhysRevX.4.031043,10.1038/ncomms8654,Houck2016}, optical fibers, waveguides, and photonic crystals \cite{Greentree_2006,PhysRevLett.100.233602,PhysRevLett.104.203603,10.1038/nature13188,Thompson1202,10.1038/ncomms4808}, small-mode-volume optical resonators \cite{10.1038/nature05586,10.1038/nature05147}, and Rydberg ensembles \cite{PhysRevLett.107.133602,10.1038/nature11361,Carr13,PhysRevLett.113.023006} all making progress towards realizing large-scale arrays of short-range coupled quantum optical building blocks.  These developments have led many researchers to revisit fundamental questions surrounding the fate of non-equilibrium quantum-optical systems in situations where, due to the importance of either dissipative or quantum fluctuations, a mean-field description is insufficient \cite{0034.4885.79.9.096001,gopal_s_nature,PhysRevA.82.043612,Lee13,PhysRevA.90.021603,PhysRevLett.110.195301,PhysRevX.6.031011,Altman15,PhysRevLett.114.040402}.  The primary goal of this paper is to elucidate the physics of a canonical many-body model made relevant by these developments---the driven-dissipative Bose Hubbard model \cite{10.1038/nphys462,PhysRevLett.103.033601,PhysRevLett.110.233601,PhysRevA.94.033801}---by bringing together many closely related ideas from the many-body physics and quantum-optics communities.  First and foremost, we aim to understand the effects of fluctuations on any phase transitions exhibited by the steady state of this model, and the extent to which these phase transitions do or do not admit an equilibrium description.  The driven-dissipative Bose-Hubbard model furnishes a minimal description of, e.g., coherently driven exciton-polariton fluids confined in coupled micro-cavities or other patterned semi-conductor devices \cite{10.1038/nature06334,PhysRevLett.112.116402,Hamel2015,Rodriguez2016}; our theoretical results thus have direct experimental implications, for example determining the universal spectral features of such systems near a dissipative steady-state phase transition.

Much of the previous work on the driven-dissipative Bose-Hubbard model has grown out of early proposals to simulate the equilibrium Bose-Hubbard model in photonic systems, either in the transient regime of very \emph{weakly} dissipative systems \cite{10.1038/nphys462,PhysRevA.76.031805}, or through clever strategies to mitigate the effects of particle loss \cite{PhysRevLett.109.053601,PhysRevB.92.174305,PhysRevX.4.031039}.  In this context, the driven-dissipative model has been considered in an attempt to understand the corruption of equilibrium physics by non-vanishing dissipation in realistic systems, and to identify qualitative signatures of equilibrium Bose-Hubbard physics---e.g. fermionization for strong interactions \cite{PhysRev.130.1605,PhysRevLett.103.033601} or the incompressibility of the zero-temperature Mott-insulating phase \cite{Fisher89,PhysRevA.90.063821,Lebreuilly2016}---that survive in steady state.  The general spirit of this approach is to start from the intuitions and expectations appropriate for the equilibrium Bose-Hubbard model, and to build outward toward an understanding of the driven-dissipative dynamics; numerous interesting connections to the equilibrium physics of the Bose-Hubbard model \cite{PhysRevLett.103.033601,PhysRevA.82.043612,PhysRevLett.108.206809}, as well as a variety of surprising and genuinely non-equilibrium effects \cite{gerace_2009,PhysRevLett.103.033601,PhysRevLett.104.113601,PhysRevLett.108.233603,PhysRevA.87.053846}, have been discovered in this manner. But there are many reasons to expect that the search for \emph{universal} features of the driven-dissipative model benefits from, and perhaps even requires, a fundamentally different approach.  For example, the ground-state and thermal phase transitions of the Bose-Hubbard model are intimately related to U(1) symmetry and the associated particle-number conservation \cite{sachdev_book}.  While the former can be preserved in a driven-dissipative context by pumping the cavities incoherently \cite{RevModPhys.85.299}, the latter remains absent \footnote{See Ref.\,\cite{0034.4885.79.9.096001} for a detailed discussion of the subtle connection between symmetries and conservation laws in driven-dissipative systems.}, calling into question whether---in the sense of phase transitions and universality---\emph{any} properties of the Bose-Hubbard model can survive the presence of driving and dissipation \cite{PhysRevLett.114.040402}.

\begin{figure}[!t]
\includegraphics[width=0.84\columnwidth]{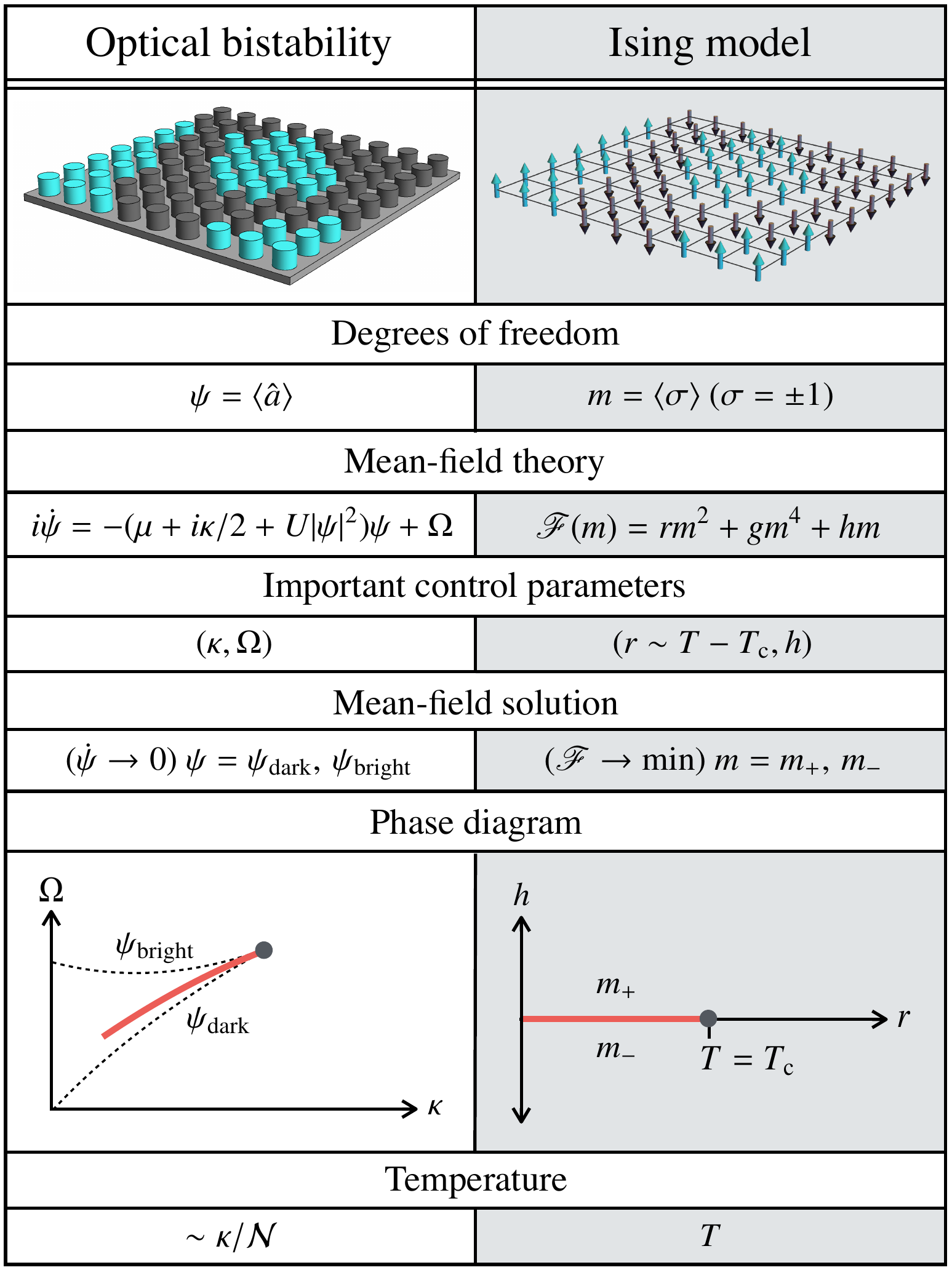}
\caption{(Color online) Summary of the correspondence between many-body optical bistability and the classical Ising model that serves as its effective equilibrium description.  The two possible magnetizations of the Ising model correspond to the bright and dark mean-field steady states of the optically bistable cavities. All parameters and variables are defined in the manuscript. In the bottom row, $\mathcal{N}$ is a parameter controlling the density scale, i.e. $|\psi|^2\sim\mathcal{N}$. Hence the low temperature limit of the Ising model corresponds to a semi-classical (large density) limit of optical bistability.}
\label{fig:schematic}
\end{figure}

Here, we instead pursue an understanding of the driven-dissipative Bose-Hubbard model from the ground up, starting from the well understood non-equilibrium physics of a single cavity, namely optical bistability \cite{Drummond:1980fg}, and adopting a functional integral formalism that is well suited to extending the essential single-cavity physics to a many-body setting \cite{PhysRevA.82.043612,PhysRevA.87.023831,PhysRevLett.110.195301,0034.4885.79.9.096001}.  As we will show, universal properties of the steady state bear essentially no resemblance to the equilibrium physics of the Bose-Hubbard model \cite{Fisher89}, but neither do they retain the fundamentally non-equilibrium character of optical bistability.  Instead, the steady state of the driven-dissipative Bose-Hubbard model admits an emergent equilibrium description in terms of a finite-temperature classical Ising model \cite{PhysRevB.93.014307}.  Specifically (see \fref{fig:schematic}), two collective mean-field steady states are inherited from the optical bistability of the individual cavities; they play the role of the two local minima in the Ising model's mean-field free energy, while dissipation (i.e.\,vacuum fluctuations) plays the role of thermal fluctuations, setting the effective temperature.  By connecting two canonical and minimal models of many-body physics---one a cornerstone of non-equilibrium quantum optics and one a cornerstone of traditional equilibrium many-body physics---this paper provides a particularly simple and concrete example of the way in which equilibrium can emerge very naturally from an \emph{a priori} non-equilibrium many-body problem, even when mean-field theory fails.  We note that our conclusions build on and justify a preliminary analysis reported in Ref.\,\cite{PhysRevB.93.014307}, where a renormalization-group argument in favor of the Ising universality class was advanced.  In the present work, we identify a small parameter $1/\mathcal{N}$ related to the inverse ``system size'' (in the sense of the system-size expansion often employed in quantum-optics studies of systems with only a few degrees of freedom), which controls the overall scale of fluctuations, and thus the effective temperature.  In the limit of weak fluctuations, the qualitative predictions of Ref.\,\cite{PhysRevB.93.014307} can be justified, made quantitative, and verified numerically.  In this way, we not only identify the model with respect to which an effective thermal description emerges, but also semi-quantitatively obtain the phase boundaries, effective temperature, and near-critical dynamics in terms of microscopic parameters.

Before proceeding, we caution that the emergence of an effective equilibrium description as described in this paper, while potentially reasonably generic, should not be taken for granted; other more genuinely non-equilibrium situations can and do arise in other models \cite{Lee13,PhysRevLett.110.195301,Chan15,Altman15,PhysRevLett.116.245701,PhysRevLett.116.070407}.  Ultimately, the goal of this paper is not only to provide a detailed view into the mechanisms by which thermal equilibrium can emerge from the microscopically non-equilibrium setting of many-body cavity-QED, nor by any means to insist that one must emerge, but also to clarify some of the techniques required to understand more unusual behaviors made possible by the strong-coupling regime of quantum optics. It is natural to expect that analyses of prototypical situations in which equilibrium \emph{does} emerge will play an important role in anticipating more exotic situations in which it does \emph{not}.

After presenting the model and reviewing the well-known solution of the single-cavity case in Sec.\,\ref{sec:model}, our general strategy for the many-body problem will be laid out in Sec.\,\ref{sec:formalism}. Our approach is based on a well-established exact mapping of the master equation onto a functional integral.  Although it imposes some additional notational burden, the functional integral formalism has the virtue of (1) being a convenient starting point for the identification of approximation schemes, including controlled strategies for going beyond mean-field theory \cite{PhysRevA.87.023831}, and (2) enabling powerful techniques such as the renormalization group to be applied \cite{0034.4885.79.9.096001,PhysRevB.93.014307}.  Here, we use the path integral to quickly identify an exactly solvable limit of the problem, around which a semi-classical expansion (related to the system-size expansion of quantum optics) can be made.  To leading non-trivial order in this expansion, we obtain a quantitatively accurate mapping of the many-body quantum master equation onto classical non-equilibrium Langevin equations, with a small parameter controlling the strength of the noise.  In Sec.\,\ref{sec:mft} we analyze the mean-field equations of motion near the mean-field critical point, which possess an emergent $Z_2$ symmetry in the spirit of Ref.\,\cite{PhysRevLett.113.210401}.  We show how the complex order parameter decomposes into two real components, one of which stays massive at the critical point and one of which does not.  By adiabatically eliminating the massive component, we arrive at a time-dependent Landau-Ginsburg equation for a scalar field, which supports two different homogeneous solutions within the bistable region.  Near the critical point and inside the bistable region, we are able to analytically obtain the profile and velocity of domain walls separating domains of these two different phases, and the vanishing of the domain wall velocity gives a zeroth-order approximation to the location of a true (first-order) phase transition in more than one spatial dimension.  In Sec.\,\ref{sec:langevin} we consider the effects of fluctuations in both one and two spatial dimensions by (a) arguing that---near the critical point and for weak noise---the non-linear Langevin equations become equivalent to model A of the Hohenberg and Halperin classification, and (b) solving the non-equilibrium Langevin equations numerically, which is valid even away from the critical point.  As our earlier analysis would suggest, the numerical results are qualitatively consistent with the expected equilibrium physics of a classical Ising model in a longitudinal field. In one dimension, domains are seeded by fluctuations, and the dynamics of their unbound domain walls smooths the mean-field transition into a crossover.  In 2D the domains walls exhibit a surface tension, enabling a line of true first-order phase transitions terminating at a critical point.

\section{Model\label{sec:model}}

The model we consider can arise in a variety of contexts, but for concreteness we consider either a 1D chain or 2D rectangular array of semiconductor microcavities supporting exciton-polaritons (see e.g. Ref.\,\cite{Rodriguez2016}).  We assume that the onsite energies of exciton-polaritons are spatially uniform and equal to $\omega_c$, and that the cavities are driven coherently and in phase by a laser with frequency $\omega_L$.  Upon making a unitary transformation to remove the time-dependence of the driving, we obtain the Hamiltonian
\begin{align}
\label{eq:hamiltonian}
\hat{H}&=-J\sum_{\langle j,k\rangle} \hat{a}^{\dagger}_j\hat{a}^{\phantom\dagger}_k-\delta\sum_{j} \hat{a}^{\dagger}_j\hat{a}^{\phantom\dagger}_j+\frac{U}{2}\sum_j\hat{a}_j^{\dagger}\hat{a}_j^{\dagger}\hat{a}^{\phantom\dagger}_j\hat{a}^{\phantom\dagger}_j\nonumber\\
&+\Omega\sum_{j}(\hat{a}^{\phantom\dagger}_j+\hat{a}_j^{\dagger}).
\end{align}
Here, $\hat{a}^{\dagger}_j(\hat{a}^{\phantom\dagger}_j)$ creates(anihilates) an exciton-polariton in the $j$th cavity, $J$ parametrizes the strength of a coherent coupling of exciton-polaritons between neighboring cavities, $\delta=\omega_{L}-\omega_c$ is the detuning of the laser from cavity resonance, $U$ sets the two-body interaction energy for exciton-polaritons confined in the same cavity, and $\Omega$ is the amplitude of the coherent driving.  The notation $\langle j,k\rangle$ implies that the sum should be taken over all nearest-neighbor pairs of sites $j$ and $k$. The driving is necessary to stabilize a non-trivial steady state in the presence of particle loss out of the cavities at a rate $\kappa$.  If the loss of exciton-polaritons is treated in the Born-Markov approximation, the dynamics of the combined unitary evolution under $\hat{H}$ and loss is described by a Markovian master equation \cite{book_carmichael_1},
\begin{align}
\label{eq:meq}
\frac{d \hat{\rho}}{dt}&=-i[\hat{H},\hat{\rho}]+\frac{\kappa}{2}\sum_{j}\big(2\hat{a}^{\phantom\dagger}_j\hat{\rho} \hat{a}^{\dagger}_j-\hat{\rho}\hat{a}^{\dagger}_j\hat{a}^{\phantom\dagger}_j-\hat{a}^{\dagger}_j\hat{a}^{\phantom\dagger}_j\hat{\rho}\big).
\end{align}
More generally, Eqs. (\ref{eq:hamiltonian},\ref{eq:meq}) provide a natural (though certainly not unique) generalization of the Bose-Hubbard model to the driven ($\Omega$) and dissipative ($\kappa$) setting of quantum optics.

\begin{figure}[!t]
\includegraphics[width=0.95\columnwidth]{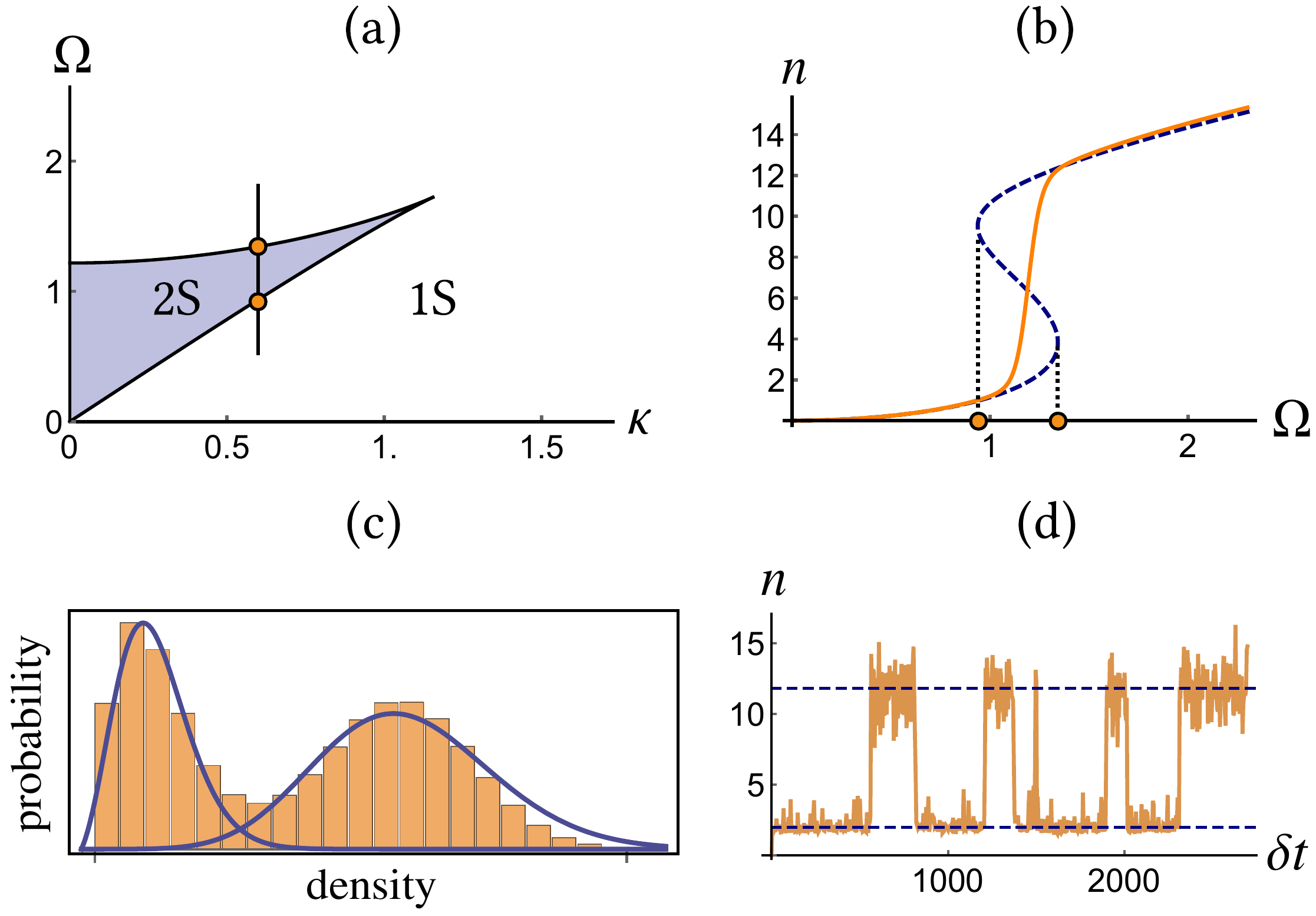}
\caption{(Color online) (a) The mean-field phase diagram for a single cavity is divided into regions that support either one or two dynamically stable solutions.  For this plot and those that follow, all parameter values are given in units of $\delta$, and $U=0.1$. (b) ($\kappa=0.6$). A cut through the bistable region of the mean-field phase diagram, showing both the mean-field (blue dashed) and exact (orange solid) solution for the density. (c) ($\kappa=0.6$ and $\Omega=1.2$). Full counting statistics of the exact steady-state density matrix, together with that of both mean-field solutions (i.e. coherent-state distributions, with the relative normalization used as a fitting parameter). (d) ($\kappa=0.6$ and $\Omega=1.2$). One trajectory obtained from a quantum trajectories simulation of \eref{eq:meq} for a single cavity, showing switching between two mean-field like states; the densities associated with the two dynamically stable mean-field solution are shown as dashed lines.}
\label{fig:mf_phase_diagram}
\end{figure}

The case of a single cavity has been thoroughly studied in the quantum optics literature, where it serves as a minimal model for dispersive optical bistability \cite{PhysRevA.17.335,Drummond:1980fg}.  A mean-field description of the problem can be obtained by writing down the equation of motion for $\psi\equiv\langle \hat{a}\rangle$, and assuming that expectation values of normal-ordered operator products factorize (i.e. making the replacement $\langle\hat{a}^{\dagger}\hat{a}\hat{a}\rangle\rightarrow|\psi|^2\psi$), giving
\begin{align}
\label{eq:mft_1}
i\dot{\psi}=-(\delta+i\kappa/2)\psi+U|\psi|^2\psi+\Omega.
\end{align}
The steady-state equation $\dot{\psi}=0$ can be recast as a cubic equation for the mean-field density $n=|\psi|^2$,
\begin{align}
\label{eq:mft_2}
n\big((\delta-Un)^2+\kappa^2/4\big)=\Omega^2.
\end{align}
This equation has either one or two solutions that are dynamically stable to small perturbations, leading to the mean-field phase diagram shown in Fig.\,\ref{fig:mf_phase_diagram}a.

For a single cavity, the full quantum solution of the master equation can be obtained efficiently in a variety of ways, for example by direct numerical integration of \eref{eq:meq} within a truncated Hilbert space.  Steady-state expectation values can even be obtained analytically by mapping the single-cavity version of \eref{eq:meq} onto phase-space equations in the complex-P representation \cite{Drummond:1980gf,Drummond:1980fg}.  Reference \cite{Drummond:1980fg} provides a comprehensive discussion of the solution, and here we simply summarize its main features, focusing primarily on the relationship between the exact and mean-field solutions.  While the mean-field equations of motion can support two dynamically stable steady states, the exact steady-state density matrix of \eref{eq:meq} is always unique, as are all observables calculated from it (for example, see \fref{fig:mf_phase_diagram}b).  While there are never two truly stable steady states, two important signatures of mean-field bistability do survive in the limit of large cavity occupancy: (1) The full-counting statistics of the exact solution exhibits a bimodal structure within the parameter regime yielding mean-field bistability, with the probability of observing different photon numbers clustering around the two mean-field stable values of the density (\fref{fig:mf_phase_diagram}c).  Outside of the bistable region, this bimodality disappears and the exact counting statistics becomes similar to that corresponding to the one stable mean-field solution; in this sense, the exact solution interpolates between the two mean-field steady states within the bistable region. (2) If the system is initialized in one of the mean-field steady states, it will only explore the phase space in the vicinity of that solution on the natural time scales of the problem (i.e. those associated with energy scales appearing explicitly in the master equation), and will only sample the phase space in the vicinity of the other mean-field solution on much longer time scales (\fref{fig:mf_phase_diagram}d).  This slow time-scale for switching between mean-field-like steady states is associated with a small gap of the exact quantum Liouvillian, which vanishes inside the bistable region in the limit of large photon occupancy \cite{PhysRevA.43.6194,PhysRevA.94.033801,ciuti_arxiv_1}. In this limit, which plays a role analogous to the ``thermodynamic limit'' of a spatially extended system \footnote{The large-density limit on a single site is closely related to the thermodynamic limit (large system-size limit at fixed density) of a model with infinite-range interactions, for which the interaction energy is also proportional to the square of the number of particles.}, mean-field bistability can therefore be identified with the existence of a (zero-dimensional) dissipative phase transition.

\section{Path integral formulation and the system size expansion\label{sec:formalism}}

To analyze the steady-state behavior of many coupled cavities, it is convenient to recast the master equation in terms of an equivalent functional integral \cite{PhysRevA.80.033624},
\begin{align}
\label{eq:partition_function}
\mathcal{Z}=\int\mathscr{D}\psi(t)\mathscr{D}\varphi(t)\mathcal{W}(\psi_0,t_0)e^{i\mathcal{S}},
\end{align}
with action
\begin{align}
\label{eq:action}
\mathcal{S}=2i&\sum_{j}\int_{t_0}^\infty dt\big(\bar{\varphi}\partial_t\psi-\varphi\partial_t\bar{\psi}\big)\\
-&\sum_{j}\int_{t_0}^{\infty}dt\big(\mathcal{H}_{\rm w}(\psi+\varphi)-\mathcal{H}_{\rm w}(\psi-\varphi)\big)\nonumber\\
+i\kappa&\sum_{j}\int_{t_0}^{\infty} dt\,\big(2\bar{\varphi}\varphi-\varphi\bar{\psi}+\bar{\varphi}\psi\big). \nonumber
\end{align}
The functional integral in \eref{eq:partition_function} is over all unconstrained paths for the variables $\psi_j(t)$ and $\varphi_j(t)$, the spatial/temporal dependence of which has been suppressed in Eqs.\,(\ref{eq:partition_function}) and (\ref{eq:action}) \footnote{Many of the usual caveats about continuum-time notation for path integrals apply here, and the interested reader should consult \aref{sec:appendix_A} for a proper discrete-time regularization of $\mathcal{Z}$.}.  The factor $\mathcal{W}(\psi_0,t_0)$ in \eref{eq:partition_function} is the Wigner function at the initial time $t_0$, with $\psi_0$ being shorthand for the set of field variables $\psi_j(0)$ at the initial time.  In this paper we will only concern ourselves with steady state properties, and so we can safely set $t_0=-\infty$ in the integral limits of \eref{eq:action} and exclude the dependence on $\mathcal{W}(\psi_0,-\infty)$ from $\mathcal{Z}$.  From here forward the integral limits of $\pm\infty$ in the action will be implied and not shown.  The classical Hamiltonian $\mathcal{H}_{\rm w}$ in \eref{eq:action} is the Weyl-symbol of the Hamiltonian in \eref{eq:hamiltonian} \cite{Polkovnikov20101790,PhysRevA.68.053604}; this should be contrasted with the usual appearance of the Q-symbol in the Keldysh functional integral (see \aref{sec:appendix_A} for a derivation of this functional integral and further discussion of the emergence of Weyl-ordering).

The functional integral in \eref{eq:partition_function} is suited to calculating products of operators that are ordered along a closed-time (or Keldysh) contour \cite{doi:10.1080/00018730902850504,book_kamenev}.  In particular, writing the so-called classical ($\psi$) and quantum ($\varphi$) fields as $ (\psi^1,\psi^2)\equiv(\bar{\psi},\psi)$ and $ (\varphi^1,\varphi^2)\equiv(\bar{\varphi},\varphi)$, and writing creation/anihilation operators as $(\hat{a}^1,\hat{a}^2)\equiv(\hat{a}^{\dagger},\hat{a})$, Keldysh-ordered correlation functions can be computed as (restoring spatial/temporal indices and defining $\mu=1,2$)
\begin{align}
\label{eq:gen_op}
\big\langle\mathcal{T}_K\big(\dots\hat{a}^{\mu}_j(t^{\pm})\dots\big)\big\rangle=\big\langle\dots\big(\psi^{\mu}_{j}(t)\pm\varphi^{\mu}_j(t)\big)\dots\big\rangle_{\mathcal{Z}}.
\end{align}
On the left-hand-side of \eref{eq:gen_op} the expectation value is taken with respect to the initial density matrix, and the operators evolve in the Heisenberg picture \footnote{Note that the Heisenberg picture is to be interpreted with respect to the Hamiltonian describing the system-bath pair \emph{before} the adiabatic elimination of the bath, which we assume proceeds within the Born-Markov approximation.}.  The symbol $\mathcal{T}_{K}$ time-orders all operators whose time arguments have a `$+$' superscript, and anti-time orders those with time arguments that have a `$-$' superscript, placing all of the latter to the left of all of the former.  On the right-hand side of \eref{eq:gen_op} the expectation value is taken with respect to the functional integral $\mathcal{Z}$, i.e. it is computed by inserting the relevant fields into the integrand of \eref{eq:partition_function} ($\mathcal{Z}$ is normalized to unity by construction, as it expresses the trace of the density matrix).  In the calculations that follow, we will exploit a semi-classical limit in which the quantum field is, in a sense to be made precise, parametrically smaller than the classical field; thus we will be primarily interested in correlations of the classical field alone, which can be converted into operator expectation values by inverting \eref{eq:gen_op},
\begin{align}
\label{eq:classical_correlations}
\big\langle\psi_{j_1}^{\mu_1}(t_1)\!\dots\!\psi_{j_n}^{\mu_n}(t_n)\big\rangle_{\!\mathcal{Z}^{\phantom K}}\!\!\!\!=\frac{1}{2^n}\!\!\sum_{\sigma=\pm}\!\big\langle\mathcal{T}_K\big(\hat{a}^{\mu_1}_{j_1}(t_1^{\sigma_1})\!\dots\!\hat{a}^{\mu_n}_{j_n}(t_n^{\sigma_n})\big)\big\rangle.
\end{align}
While it may appear that such correlation functions can be discontinuous at coinciding times due to the associated change of operator ordering on the right-hand-side of \eref{eq:classical_correlations}, it is straightforward to show that this is not actually the case.  Instead, when the times approach each other ($t_1,\dots,t_n\rightarrow t$), the limit of an arbitrary $n$-point correlation function of the classical field $\psi$ smoothly approaches the equal-time value
\begin{align}
\label{eq:psi_eq_time}
\big\langle\psi_{j_1}^{\mu_1}(t)\dots\psi_{j_n}^{\mu_n}(t)\big\rangle_{\mathcal{Z}}=\big\langle \big(\hat{a}^{\mu_1}_{j_1}(t)\dots\hat{a}^{\mu_n}_{j_n}(t)\big)_{\rm s}\big\rangle,
\end{align}
where $(\dots)_{\rm s}$ symmetrizes (i.e. Weyl-orders) products of creation and annihilation operators \cite{book_carmichael_1}.  In other words, equal-time correlation functions of the classical field reproduce the average of Weyl-ordered operator products.  For example, the density can be computed from the two-point correlation function
\begin{align}
\langle \bar{\psi}_j(t)\psi_j(t)\rangle_{\mathcal{Z}}=\big\langle\big(\hat{a}_j^{\dagger}(t)\hat{a}_j^{\phantom\dagger}(t)\big)_{\rm s}\big\rangle&=\frac{1}{2}\langle\hat{a}_j^{\dagger}(t)\hat{a}_j^{\phantom\dagger}(t)+\hat{a}_j^{\phantom\dagger}(t)\hat{a}_j^{\dagger}(t)\rangle\nonumber\\
&=\langle\hat{n}(t)\rangle+\frac{1}{2}.
\end{align}

The Weyl-symbol of $\hat{H}$ is given by (ignoring additive constants, which do not affect correlation functions)
\begin{align}
\label{eq:weyl_hamiltonian}
\mathcal{H}_{\rm w}(\alpha)\!=\!\sum_{j}\Big(-\bar{\alpha}_j(J\nabla^2+\mu+U)\alpha_j +\frac{U}{2}|\alpha_j|^4+\Omega(\bar{\alpha}_j+\alpha_j)\Big).
\end{align}
Here $\nabla^2\alpha_j\equiv -z\alpha_j+\sum_{\langle k,j\rangle}\alpha_k$ is the discrete Laplacian, the lattice coordination number $z=2D$ in $D$ dimensions, and $\mu=\delta+zJ$.  Inserting \eref{eq:weyl_hamiltonian} into \eref{eq:action} yields the action
\begin{align}
\label{eq:source_free_action}
\mathcal{S}=&\,2\!\sum_{j}\!\int \!\! dt\,\bar{\varphi}\Big(i \partial_t\psi+(J\nabla^2+\mu+i\frac{\kappa}{2})\psi-\Omega-U|\psi|^2\psi\Big)+{\rm c.c.}\nonumber\\
&+2i\kappa\!\sum_j\!\int \!\!dt\,\bar{\varphi}\varphi+2U\!\sum_j\!\int\!\! dt\,(\bar{\varphi}\varphi+1)(\psi\bar{\varphi}+\bar{\psi}\varphi).
\end{align}

Because the path integral is Gaussian for $U=0$, and because optical bistability at the mean-field level can occur at arbitrarily small values of $U$, one might hope that some aspects of the relevant physics can be captured by doing perturbation theory in $U$.  However, this is not the case; while optical bistability can indeed occur for small $U$, it always occurs when the typical interaction energy per particle, $U|\psi|^2$, is comparable to the other energy scales of the problem.  Nevertheless, the action can still be organized around a small parameter that enables a controlled approximation.  To this end, we define rescaled fields and parameters
\begin{align}
\Phi\equiv\varphi\sqrt{\mathcal{N}},~~~\Psi\equiv \psi/\sqrt{\mathcal{N}},~~~\omega\equiv\Omega/\sqrt{\mathcal{N}},~~~u\equiv U\mathcal{N},\end{align}
in terms of which the action can be rewritten
\begin{align}
\label{eq:source_free_action_rescaled}
\mathcal{S}=&\,2\sum_j\!\!\int \!\!dt \,\bar{\Phi}\Big(i \partial_t\Psi+(J\nabla^2+\mu+i\frac{\kappa}{2})\Psi-\omega-u|\Psi|^2\Psi\Big)\!+{\rm c.c.}\nonumber\\
+&\frac{1}{\mathcal{N}}\!\sum_j\!\!\int \!\!dt\,(2i\kappa\bar{\Phi}\Phi)+\frac{1}{\mathcal{N}^2}\!\sum_j\!\!\int \!\!dt\,2u(\bar{\Phi}\Phi+\mathcal{N})(\Psi\bar{\Phi}+\bar{\Psi}\Phi).
\end{align}
The dimensionless parameter $\mathcal{N}$ implicitly identifies a one-dimensional family of actions at fixed values of $\omega$, $u$, $\kappa$, $\mu$, and $J$, one limit of which (large $\mathcal{N}$) will be shown to admit a tractable analysis.  Note that the limit $\mathcal{N}\rightarrow\infty$ at fixed $u$ and $\omega$ is not the same thing as the limit $U\rightarrow 0$, even though the coupling $U$ does become small in this limit.  Rather, increasing $\mathcal{N}$ amounts to increasing the drive strength $\Omega$ while simultaneously decreasing the coupling $U$ in such a way that the typical interaction energy per particle, $U|\psi|^2$, remains constant.

To see this, we first evaluate the path integral in the limit $\mathcal{N}\rightarrow\infty$, in which only the first term in the action survives. The functional integral over $\Phi$ can be carried out and yields a functional $\delta$-function of the term inside parentheses, thereby enforcing the mean-field equation of motion
\begin{align}
\label{eq:mft_lattice}
i\partial_t\Psi=-J\nabla^2\Psi-(\mu+i\kappa/2)\Psi+\omega+u|\Psi|^2\Psi.
\end{align}
Note that at this level of approximation, varying $\mathcal{N}$ with $\omega$ and $u$ held fixed leaves the equation of motion for $\Psi$ invariant.  Therefore, as discussed above, the interaction energy per photon $U|\psi|^2=u|\Psi|^2$ stays fixed. The only consequence is that the actually density, $|\psi|^2=\mathcal{N}|\Psi|^2$, is enhanced by a factor of $\mathcal{N}$, which therefore sets the overall density scale \cite{ciuti_arxiv_1}.

For $\mathcal{N}$ large but finite, the first term on the second line of \eref{eq:source_free_action_rescaled} suppresses contributions to the path integral unless $\Phi\lesssim\mathcal{N}^{1/2}$. The final term can therefore be estimated as $\Phi^3\mathcal{N}^{-2}+\Phi\mathcal{N}^{-1}\lesssim\mathcal{N}^{-1/2}$, and can be safely ignored in the large $\mathcal{N}$ limit.  At this level of approximation, the path integral can no longer be solved exactly, but it can be mapped onto stochastic classical equations by standard techniques.  Decoupling the term that is quadratic in $\Phi$ with a Hubbard-Stratonovich transformation,
\begin{align}
e^{-(2\kappa/\mathcal{N})|\Phi|^2}=\frac{2\mathcal{N}}{\kappa\pi}\int d^2\zeta e^{2i(\Phi\bar{\zeta}+\bar{\Phi}\zeta)}e^{-2\mathcal{N}|\zeta|^2/\kappa},\nonumber
\end{align}
the action again becomes linear in $\Phi$.  This time, for fixed $\zeta$, the functional integral over $\Phi$ enforces enforces the equation of motion
\begin{align}
\label{eq:sGPE}
i\partial_t\Psi=-J\nabla^2\Psi-(\mu+i\kappa/2)\Psi+\omega+u|\Psi|^2\Psi+\zeta.
\end{align}
The remaining functional integral over $\zeta$ with a gaussian weight $\exp(-2\mathcal{N}|\zeta|^2/\kappa)$ indicates that we should interpret \eref{eq:sGPE} as a stochastic differential equation, with $\zeta$ being complex, Gaussian white noise of variance (restoring spatial and temporal indices)
\begin{align}
\overline{\bar{\zeta}_j(t_1)\zeta_k(t_2)}=\frac{1}{\mathcal{N}}\frac{\kappa}{2}\delta_{j,k}\delta(t_1-t_2).
\end{align}
Hence the dynamics of the rescaled classical field $\Psi$, to this order in $1/\mathcal{N}$, is governed by a stochastic and dissipative Gross-Pitaevskii equation with parametrically weak noise \footnote{Equation \eqref{eq:sGPE} is closely related to the truncated-Wigner approximation, though there is a subtle difference; the truncated-Wigner approximation would result in an interaction term $u(|\Psi|^2-1/\mathcal{N})\Psi$ rather than $u|\Psi|^2\Psi$ owing to the Weyl-ordering of the Hamiltonian.  The additional term, however, should not be retained when making a consistent expansion in $1/\mathcal{N}$.}.  Though \eref{eq:sGPE} can also be derived using more standard phase space techniques, the path integral approach has the advantage that one can assess the consequences of the approximations that lead to \eref{eq:sGPE} within the framework of the renormalization group.  In particular, as discussed in Ref.\,\cite{PhysRevB.93.014307}, \eref{eq:sGPE} should reproduce the correct critical exponents for the phase transition exhibited by the exact steady state of \eref{eq:meq}.  Unlike in Ref.\,\cite{PhysRevB.93.014307}, however, here we have explicitly identified a limit (large $\mathcal{N}$) in which \eref{eq:sGPE} yields asymptotically exact results for microscopic observables, and thus can be used to make quantitative predictions about the behavior of correlation functions at the lattice-scale (rather than just qualitative predictions about their long-distance asymptotics).  Moreover, the existence of this limit furnishes a more formal justification for the type of perturbative renormalization-group analysis carried out in Ref.\,\cite{PhysRevB.93.014307}.

It is important to realize that, even for a single cavity, Eqs.\,\eqref{eq:mft_lattice} and \eqref{eq:sGPE} must be interpreted with some care in order to correctly extract steady-state properties in the large-$\mathcal{N}$ limit.  The reason is simply that, returning to the full path integral, the limits $\mathcal{N}\rightarrow\infty$ and $t\rightarrow\infty$ do not commute when the parameters are tuned to be inside the mean-field bistable region.  Indeed, if we take the limit $\mathcal{N}\rightarrow\infty$ first, \eref{eq:mft_lattice} is exact at all times, and we are lead to conclude (on the same basis as the analysis in \sref{sec:model}) that there are two stable steady states.  If we instead take the large $t$ limit first, we would find [based on the analysis of $1/\mathcal{N}$ corrections contained in \eref{eq:sGPE}] that there is a unique steady state at any finite value of $\mathcal{N}$, which is a fluctuation-induced admixture of the two stable steady states computed by reversing the order of limits.  If we now take the large-$\mathcal{N}$ limit, one of those steady states is generally preferred over the other, in the sense that it alone determines all steady-state expectation values.  Thus we encounter a sudden (first-order) phase transition between a bright and a dark state when traversing through the mean-field bistable region.

\section{Mean-field theory\label{sec:mft}}

\begin{figure}[!t]
\includegraphics[width=0.95\columnwidth]{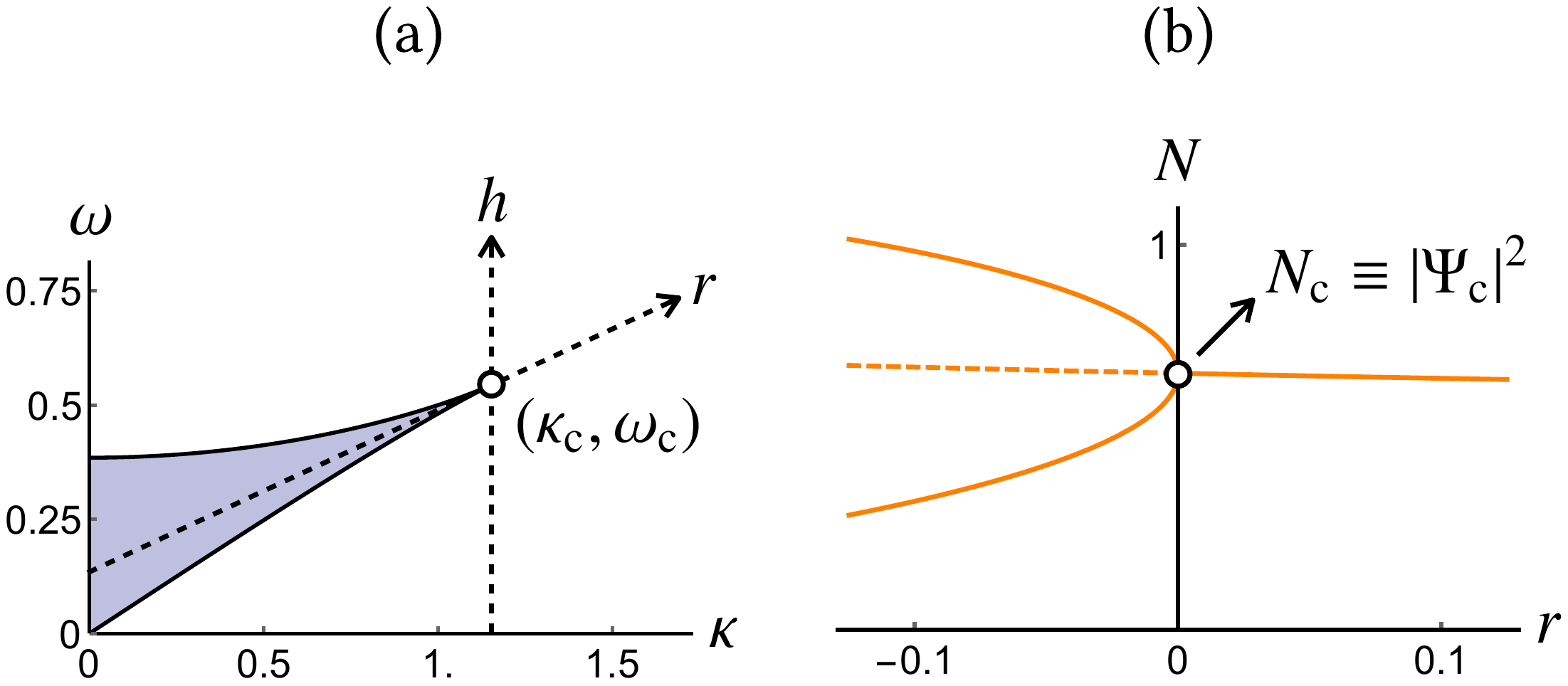}
\caption{(Color online) (a) Coordinates used to parametrize passage through the mean-field critical point and into the bistable region; $r$ and $h$ control deviations in a non-orthogonal coordinate system spanned by the dashed arrow. (b) Setting $h=0$ and scanning $r$ from positive to negative causes the order parameter to undergo a cusp bifurcation, at which the solution outside the bistable region goes unstable (dashed line) and two new dynamically stable steady states (solid lines) emerge continuously.  In both plots $u=\mu$, and energies are given as dimensionless ratios with $\mu$.}
\label{fig:mf_coordinates}
\end{figure}

From the above considerations it is clear that, at least in the large $\mathcal{N}$ limit, the steady state can be understood by solving the mean-field equations of motion in the presence of parametrically weak noise.  Thus we expect a detailed understanding of the mean-field dynamics in the absence of noise to form a useful starting point for understanding the dynamics of \eref{eq:sGPE}.  At the level of \eref{eq:mft_lattice}, and assuming that only uniform steady states exist, the steady-state phase diagram is identical to that of single-cavity optical bistability, up to the replacement of $\delta$ by $\mu$.  In terms of the rescaled field $\Psi$ and a rescaled density $N\equiv|\Psi|^2=\mathcal{N}^{-1}|\psi|^2$, \eref{eq:mft_2} becomes
\begin{align}
\label{mft_3}
N\big((\mu-uN)^2+\kappa^2/4\big)=\omega^2.
\end{align}
Straightforward analysis of \eref{mft_3} shows that upon entering the bistable region from outside of it, the additional solution does not in general emerge continuously from the existing one.  However, if one enters the bistable region through the cusp located at (see \fref{fig:mf_coordinates})
\begin{align}
\{\kappa_{\rm c},\omega_{\rm c}\}=\mu\{(4/3)^{1/2},(2/3)^{3/2}(\mu/u)^{1/2}\},
\end{align}
the two solutions do emerge continuously from a single solution, $\Psi_{\rm c}=e^{-i\pi/3}\sqrt{2\mu/3u}$ (with critical density $N_{\rm c}=|\Psi_{\rm c}|^2=2\mu/3u$).  Therefore, we can identify the cusp of the bistable region as a mean-field critical point locating a continuous phase transition from one to two steady states.

It is straightforward to show that one can only enter the bistable region through the critical point along the line $(\kappa-\kappa_{\rm c})=\sqrt{8u/\mu}(\omega-\omega_{\rm c})$.  It will be convenient in what follows to define new coordinates in the $\kappa$-$\omega$ parameter space that naturally parametrize deviations from the mean-field critical point along and away from this line,
\begin{align}
r=\frac{1}{2}(\kappa-\kappa_{\rm c}),~~~h=\frac{4}{\sqrt{3}}(\omega-\omega_{\rm c})-\sqrt{\frac{2\mu}{3u}}(\kappa-\kappa_{\rm c}).
\end{align}
Note that this parameter transformation is designed so that only $r$ varies as we enter the bistable region through the critical point---the overall normalization of $r$ and $h$ is arbitrary, and chosen to make the formulas that follow simpler.  These new variables can be visualized as parameterizing deviations from the mean-field critical point in the non-orthogonal coordinate system shown in \fref{fig:mf_coordinates}a.  For $h=0$, moving from $r>0$ to $r<0$ causes the mean-field solution outside of the bistable region to undergo a cusp bifurcation (\fref{fig:mf_coordinates}b).  For $r<0$, sweeping $h$ from negative to positive traverses the bistable region in such a way that the system goes from supporting only a low-density solution, to having coexisting low-density and high-density solutions, and then eventually to supporting only a high-density solution.  This behavior is in close analogy to that of an Ising model: If the dark and bright solutions are identified with the up/down-symmetry-related  free-energy minima, then $r$ plays the role or the reduced temperature and $h$ plays the role of a symmetry-breaking (longitudinal) field, causing one to be preferred over the other.  The remainder of \sref{sec:mft} will formalize this analogy, and in \sref{sec:langevin} we will argue that it continues to hold even when fluctuations are included.

\subsection{Near-critical dynamics}

We are particularly interested in the effects of fluctuations in the vicinity of the mean-field critical point, which requires that we first understand the mean-field response when (a) the parameters are tuned close to the mean-field critical point (both $r$ and $h$ are small) and (b) the order parameter $\Psi$ is perturbed weakly from its steady-state value.  First working directly at the critical point ($r=h=0$, with steady-state solution $\Psi=\Psi_{\rm c}$) and assuming that $\Psi=\Psi_{\rm c}+\delta\Psi$ is uniform and close to the critical value, we expand \eref{eq:mft_lattice} to first order in $\delta\Psi$ to obtain
\begin{align}
\label{eq:mf_at_critical}
\partial_t\delta\Psi=-\frac{\mu}{2}\bigg(\big(\sqrt{3}+i\big)\delta\Psi+\big(\sqrt{3}-i\big)\delta\bar{\Psi}\bigg).
\end{align}
The r.h.s. of \eref{eq:mf_at_critical} is purely real, and thus only the real part of $\delta\Psi$ decays; the dynamics stops when the r.h.s vanishes, i.e. when ${\rm arg}(\delta\Psi)=\pi/2-{\rm arg}(\sqrt{3}+i)=\pi/3$.  This observation motivates the following decomposition of the complex-valued $\delta\Psi$ into two real components,
\begin{align}
\label{eq:psi_decomp}
\delta\Psi=\varrho+e^{i\pi/3}\sigma,
\end{align}
with the expectation that $\varrho$ and $\sigma$ will relax quickly and slowly, respectively, in the vicinity of the mean-field critical point.  Inserting this decomposition into \eref{eq:mft_lattice} but now keeping all orders in $\varrho$ and $\sigma$, we obtain coupled non-linear differential equations for $\varrho$ and $\sigma$ (see Appendix \ref{sec:appendix_B} for a detailed discussion).  The fast variable $\varrho$ can be adiabatically eliminated by solving $\partial_t \varrho=0$ for $\varrho$ (perturbatively in $r$ and $h$) and inserting the solution into the equation of motion for $\sigma$.  In this way, to lowest nontrivial order in $r$ we obtain
\begin{align}
\label{eq:vdot}
\partial_{t}\sigma=\frac{J}{\sqrt{3}}\nabla^2\sigma-r\sigma -\frac{u}{\sqrt{3}}\sigma^3-\frac{h}{2}.
\end{align}
Note that we have also dropped higher-order derivative terms for $\sigma$ that arise from the adiabatic elimination of $\varrho$; this omission turns out to be justified (to lowest non-trival order in $r$) near the mean-field critical point, where the fields vary slowly in space, even when $J$ is not small compared to the other energy scales in \eref{eq:vdot} (the perturbative adiabatic elimination of $\varrho$ is explained in detail in Appendix \ref{sec:appendix_B}).  Restoring spatial indices and defining parameters
\begin{align}
K=J/\sqrt{3},~~~~g=u/\sqrt{3},
\end{align}
\eref{eq:vdot} can be rewritten as
\begin{align}
\label{eq:relaxational}
\partial_{t}\sigma_j=-\frac{\partial\mathscr{H}(\sigma)}{\partial \sigma_j},
\end{align}
where the effective Hamiltonian $\mathscr{H}(\sigma)$ is defined
\begin{align}
\label{eq:effective_hamiltonian}
\mathscr{H}(\sigma)=\frac{1}{2}\sum_{j}\bigg(K|\nabla \sigma_j|^2+r \sigma_j^2 +\frac{1}{2}g\sigma_j^4+h \sigma_j\bigg).
\end{align}
Note that, as anticipated, $\mathscr{H}(\sigma)$ is precisely the energy functional defining the Landau theory of a classical Ising model, with $\sigma$ playing the role of the magnetization.  Equation \eqref{eq:relaxational} indicates that the dynamics of the slow field in the vicinity of the critical point is purely relaxational, evolving towards the minimum of the effective potential $\mathscr{H}(\sigma)$.

\subsection{Domain walls}

At the level of mean-field theory there are two truly stable homogeneous solutions within the bistable region.  However, it is clear that if we place the system in one of the two mean-field steady states, the inclusion of fluctuations will seed defects of the other steady state; whether these defects shrink or grow will depend on the dynamics of the domain wall separating them from the bulk, and will determine which of the two mean-field steady states is preferred over the other.  Thus the identification of a point in the bistable region where the mean-field velocity of a domain wall vanishes gives a first approximation to the location of the phase transition when fluctuations are included.

It is difficult to analytically extract domain-wall dynamics directly from \eref{eq:mft_lattice}, so to proceed we make three assumptions: (1) The parameters are tuned to be inside the bistable region and close to the mean-field critical point, (2) The domains are smooth, such that a continuum approximation is justified, and (3) If $D>1$, the domains are large and thus have vanishing curvature. The first assumption justifies the use of \eref{eq:relaxational} to calculate the dynamics. The second assumption requires that $J$ is large compared to the local energy scales of the problem, e.g. to the characteristic time-scale associated with dynamics in the potential part of the Hamiltonian
\begin{align}
\label{eq:u_of_v}
U(\sigma)=\frac{1}{2}\big(r \sigma^2+\frac{1}{2}g \sigma^4+h \sigma\big).
\end{align}
Note that near criticality, this only requires that $J$ is large compared to $r$ and $h$, and not that $J$ is large compared to the energy scales $\omega$ and $\gamma$.  The third assumption is made because the phase that is favored in the limit of weak fluctuations is the one in which asymptotically large defects of the opposite phase are unfavored (i.e. tend to shrink).

\begin{figure}[!t]
\includegraphics[width=1\columnwidth]{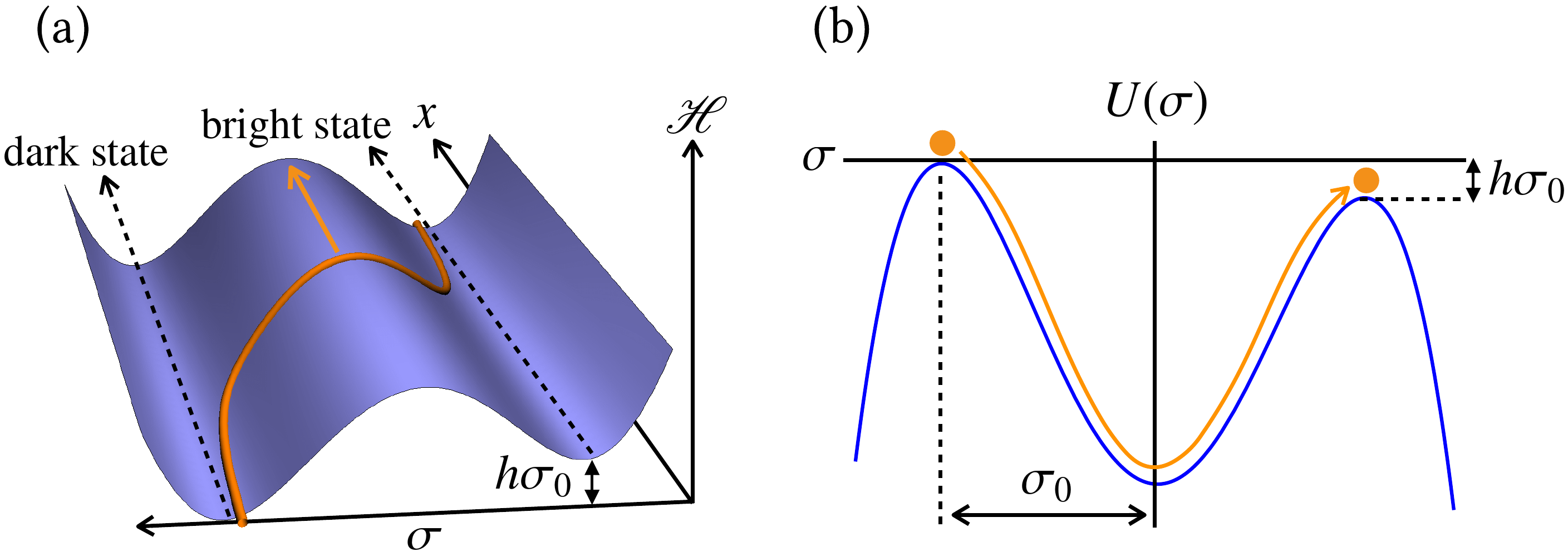}
\caption{(Color online) (a) Domain wall dynamics near the critical point.  Since the dynamics is relaxational, the domain wall moves in such a direction that the lower-energy domain increases in size.  (b) This dynamics can be mapped onto the motion of a fictitious particle in an inverted potential.}
\label{fig:domains}
\end{figure}

The dynamics of a flat domain wall are effectively one-dimensional, and can be ascertained from a one-dimensional continuum version of \eref{eq:relaxational},
\begin{align}
\label{eq:vdot_1d}
\partial_t\sigma(x,t)=K\partial_x^2\sigma(x,t)-r \sigma(x,t) -g \sigma(x,t)^3-\frac{h}{2}.
\end{align}
When $h=0$, the symmetry of \eref{eq:vdot_1d} under inversions $\sigma\rightarrow -\sigma$ implies that domain walls must be stationary; both uniform phases have the same effective potential and relaxational dynamics cannot prefer one over the other.  Therefore, the line $h=0$ provides a first approximation to the dividing line between parts of the bistable region in which the bright phase is more stable and parts in which the dark phase is more stable.

\begin{figure}[!t]
\includegraphics[width=0.95\columnwidth]{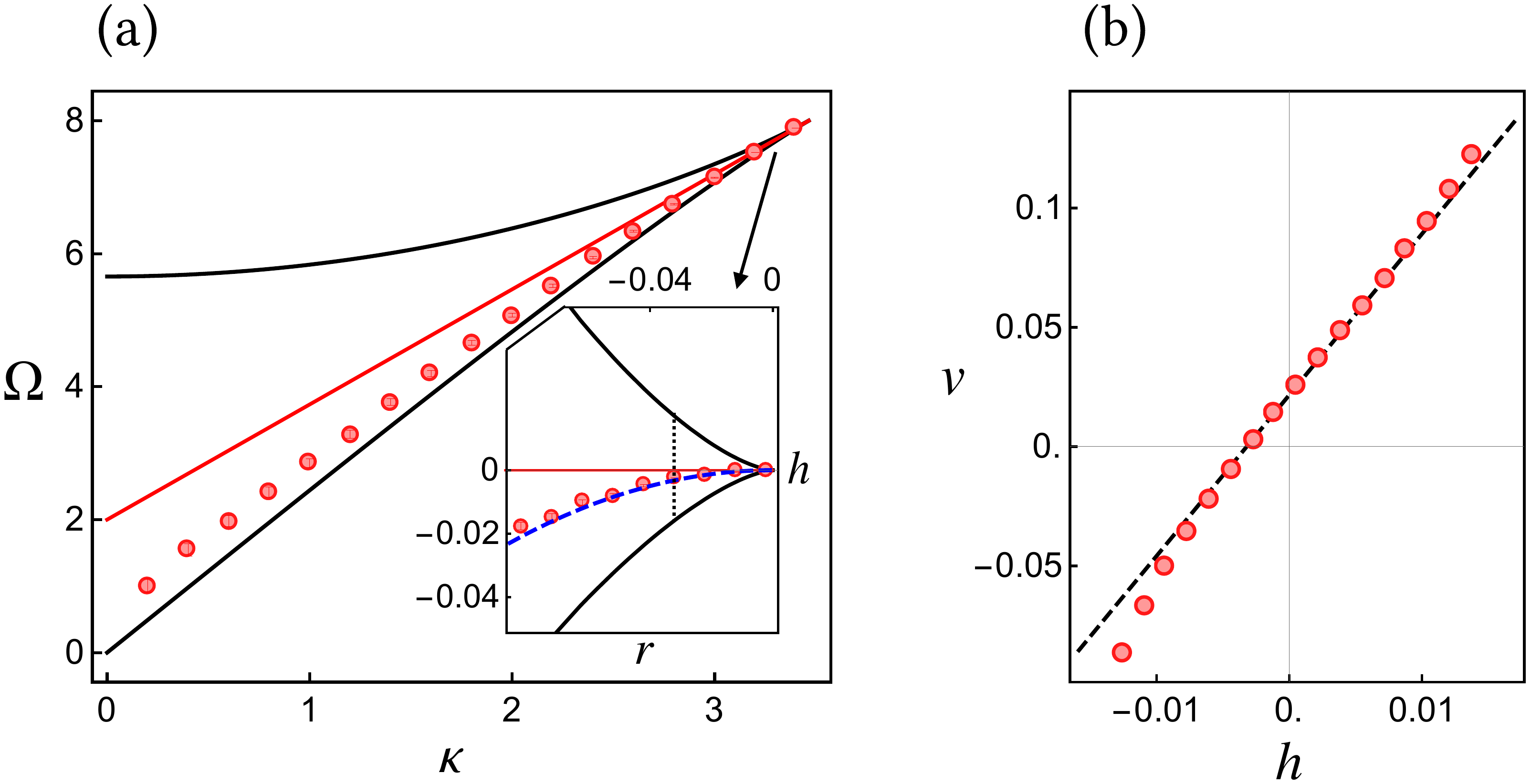}
\caption{(Color online) (a) Numerically determined location of the domain-wall velocity zeros (red disks), together with the analytical estimate of the zero-velocity line, $h=0$, valid near the critical point (red line). The size of the disks reflects the largest expected uncertainty in the numerical determination of these points.  Inset: exploded view of the main plot near the critical point, now in terms of the parameters $h$ and $r$.  The dashed blue line is an improved estimate of the zero-velocity line obtained by extending \eref{eq:vdot} to next leading (quadratic) order in $r$.  (b) Numerically extracted domain wall velocity (red disks) as a function of $h$ [taken along the black-dotted line shown in the inset of (a)], compared with the estimate in \eref{eq:v_approx_text} (black-dashed line).  Note that the black-dashed line does not vanish at $h=0$.  While its slope is taken from \eref{eq:v_approx_text}, it has been shifted by an amount that we determine by extending \eref{eq:vdot} to next-to-leading order in the deviations from the critical point [i.e. the same correction used to produce the blue-dashed line in the inset of (a)], which clearly agrees well with the numerically calculated shift of the zero-velocity point.}
\label{fig:velocity}
\end{figure}

For $h$ small but nonzero, the domain wall velocity can be estimated in the following manner \cite{LANGER1967108}.  Making a traveling wave ansatz $\sigma(x,t)=\sigma(\tau)$, with $\tau\equiv x- vt$, and denoting derivatives with respect to $\tau$ by dots, \eref{eq:vdot_1d} becomes
\begin{align}
\label{eq:particle_analogy_text}
K\ddot{\sigma}=-v\dot{\sigma}+\frac{\partial U(\sigma)}{\partial \sigma}.
\end{align}
Equation \eqref{eq:particle_analogy_text} can be interpreted as Newton's equation for a particle with position $\sigma$ and mass $K$, moving in the inverted potential $-U(\sigma)$ and subject to a linear drag with friction coefficient $v$. Inside the mean-field bistable region ($r<0$) there are two stationary solutions of \eref{eq:particle_analogy_text} associated with the two local maxima of the potential energy $-U(\sigma)$ (\fref{fig:domains}); these correspond to the two spatially uniform mean-field steady states.  To zeroth order in $h$, these solutions are located at $\sigma_{\pm}=\pm \sigma_0$, with $\sigma_0=\sqrt{|r|/g}$.  In addition to the two stationary stationary solutions, a solution can be found that interpolates from the higher local maximum to the lower one, which for $h>0$ is located at $\sigma=+\sigma_0$. The friction coefficient $v$ must be determined self-consistently such that the particle comes to rest at the lower local maximum.  Standard analysis of the solutions of \eref{eq:particle_analogy_text} based on conservation of energy yields, to first order in $h$ (see Appendix \ref{sec:appendix_C} for details),
\begin{align}
\label{eq:v_approx_text}
v\approx h\frac{3}{2}\sqrt{\frac{K g}{2r^2}}.
\end{align}
The analysis above is corroborated by brute-force numerical integration of \eref{eq:mft_lattice}.  The true zero-velocity line can be determined numerically by solving \eref{eq:mft_lattice} with a domain wall inserted at $t=0$, and agrees with the $h=0$ line near the critical point (\fref{fig:velocity}a).  Also, as shown in \fref{fig:velocity}b, \eref{eq:v_approx_text} agrees well with the numerically extracted domain-wall velocity.

\section{Fluctuations\label{sec:langevin}}

Mean-field theory suggests that the steady state of the master equation in \eref{eq:meq} undergoes an Ising-like phase transition in sufficiently high spatial dimensions.  However, in order to understand the detailed nature of this phase transition, and to determine its lower critical dimension, fluctuations must be taken into account. As discussed in \sref{sec:formalism}, for large $\mathcal{N}$ the dominant fluctuations are captured by working with the stochastic and dissipative Gross-Pitaevskii equation in \eref{eq:sGPE}, reproduced here for clarity,
\begin{align}
\label{eq:sGPE_2}
i\partial_t\Psi=-J\nabla^2\Psi-(\mu+i\kappa/2)\Psi+\omega+u|\Psi|^2\Psi+\zeta.
\end{align}
At this level of approximation, expectation values of the classical field are obtained by averaging the solution of \eref{eq:sGPE_2} over realizations of the noise $\zeta$, $\langle\dots\rangle_{\mathcal{Z}}\approx\langle\dots\rangle_{\rm sGPE}$.

Though we cannot solve \eref{eq:sGPE_2} analytically, simple arguments can be made to explain many features of the steady state quantitatively near the mean-field critical point, and qualitatively even away from it.  As before, the near critical dynamics is simplified by decomposing the field as $\Psi=\Psi_{\rm c}+(\varrho+e^{i\pi/3}\sigma)$.  Adiabatic elimination of $\varrho$ can again be performed perturbatively in $h$ and $r$; the only subtlety is that fluctuations cause $\varrho$ to undergo a lattice version of the Ornstein-Uhlenbeck process \cite{PhysRev.36.823}, which feeds back into the equation of motion for $\sigma$ as non-$\delta$-correlated noise.  However, it is straightforward to show that near the critical point the correlation-time of this additional noise is short compared to the dynamical timescales of $\sigma$, and it can be incorporated as a perturbative renormalization of the $\delta$-correlated noise acting directly on $\sigma$.  Details of the calculation are reported in Appendix \ref{sec:appendix_B}, and here we simply quote the final result,
\begin{align}
\label{eq:vdot_noise}
\partial_t\sigma_j=-\frac{\partial\mathscr{H}(\sigma)}{\partial \sigma_j}+\xi_j(t).
\end{align}
Here, $\mathscr{H}(\sigma)$ is the same energy functional given in \eref{eq:effective_hamiltonian}, and $\xi_j(t)$ is (real) Gaussian white noise with variance
\begin{align}
\overline{\xi_j(t_1)\xi_k(t_2)}=\frac{\kappa}{3\mathcal{N}}\delta_{j,k}\delta(t_1-t_2).
\end{align}

Equation \eqref{eq:vdot_noise} is a spatially discretized version of Model A in the Hohenberg-Halperin classification \cite{RevModPhys.49.435}, suggesting that the steady-state phase transition associated with optical bistability in the driven-dissipative Bose-Hubbard model is, as anticipated, in the universality class of the finite-temperature classical Ising model.  In particular, steady-state and static observables generated by the stochastic dynamics in \eref{eq:vdot_noise} can be computed with respect to the Boltzmann weight
\begin{align}
\mathscr{P}(\sigma)=\mathcal{Z}^{-1}e^{-\mathscr{H}(\sigma)/T_{\rm eff}},~~\mathcal{Z}=\int \prod_{j}d\sigma_j e^{-\mathscr{H}(\sigma)/T_{\rm eff}},
\end{align}
with an effective temperature given by
\begin{align}
T_{\rm eff}=\kappa/3\mathcal{N}.
\end{align}
The large $\mathcal{N}$ limit was designed to suppress fluctuations in the microscopic action and so, unsurprisingly, it corresponds to a low-temperature limit of the effective equilibrium description of the phase transition.  Returning to the underlying microscopic degrees of freedom, it is straightforward to see that the dynamics of this effective theory is imprinted on experimentally measurable observables.  The connection is particularly simple near the mean-field critical point.  For example, working to lowest non-trivial order in the deviations of the fields from their mean-field critical values, straightforward algebra yields the equal-time connected density-density correlation function
\begin{align}
\mathcal{C}_{jk}&=\langle\hat{n}_j(t)\hat{n}_k(t)\rangle-\langle\hat{n}_j(t)\rangle\langle \hat{n}_k(t)\rangle\nonumber\\
&\propto\langle\sigma_j(t)\sigma_k(t)\rangle_{\rm sGPE}-\langle\sigma_j(t)\rangle_{\rm sGPE}\langle\sigma_k(t)\rangle_{\rm sGPE}.
\end{align}
The critical properties of the finite-temperature Ising model should, therefore, control the critical fluctuations of the intensity of light emitted from a coherently-driven array of exciton-polariton microcavities.

Before considering what happens away from the critical point, we first briefly summarize a qualitative picture of model A dynamics and its connection to the Ising model. Suppose that the system is seeded in a locally random initial configuration: We would like to know what happens to it in steady state. At short times and for $r<0$, we expect the system to form domains of both (locally stable) phases, separated by domain walls.  In the absence of fluctuations (i.e. at $T_{\rm eff}=0$) the preferred steady state of the system can be understood by simple domain-wall dynamics; for $h\neq 0$ one phase is preferred over the other, and the system will eventually order in that phase. If fluctuations are now turned on, domains of the less favored phase will be seeded, and the consequence of these defects depends crucially on the dimensionality.  In one dimension, the domain walls enclosing these defects move independently of each other when they are sufficiently far apart, undergoing a biased random walk.  As a result, when $h\rightarrow 0$ and the dynamics becomes unbiased, defects proliferate and the system will be disordered at any finite temperature.  In two or more spatial dimensions, defects of the less favored phase will still be seeded by fluctuations, but small defects contract aggressively even as $h\rightarrow 0$ due to a surface tension.  Therefore, at least at sufficiently small temperature, as $h\rightarrow 0$ the system remains ordered in a phase that depends on whether $h$ approaches zero from below or above, indicating a first-order phase transition.

\begin{figure}[!t]
\includegraphics[width=0.9\columnwidth]{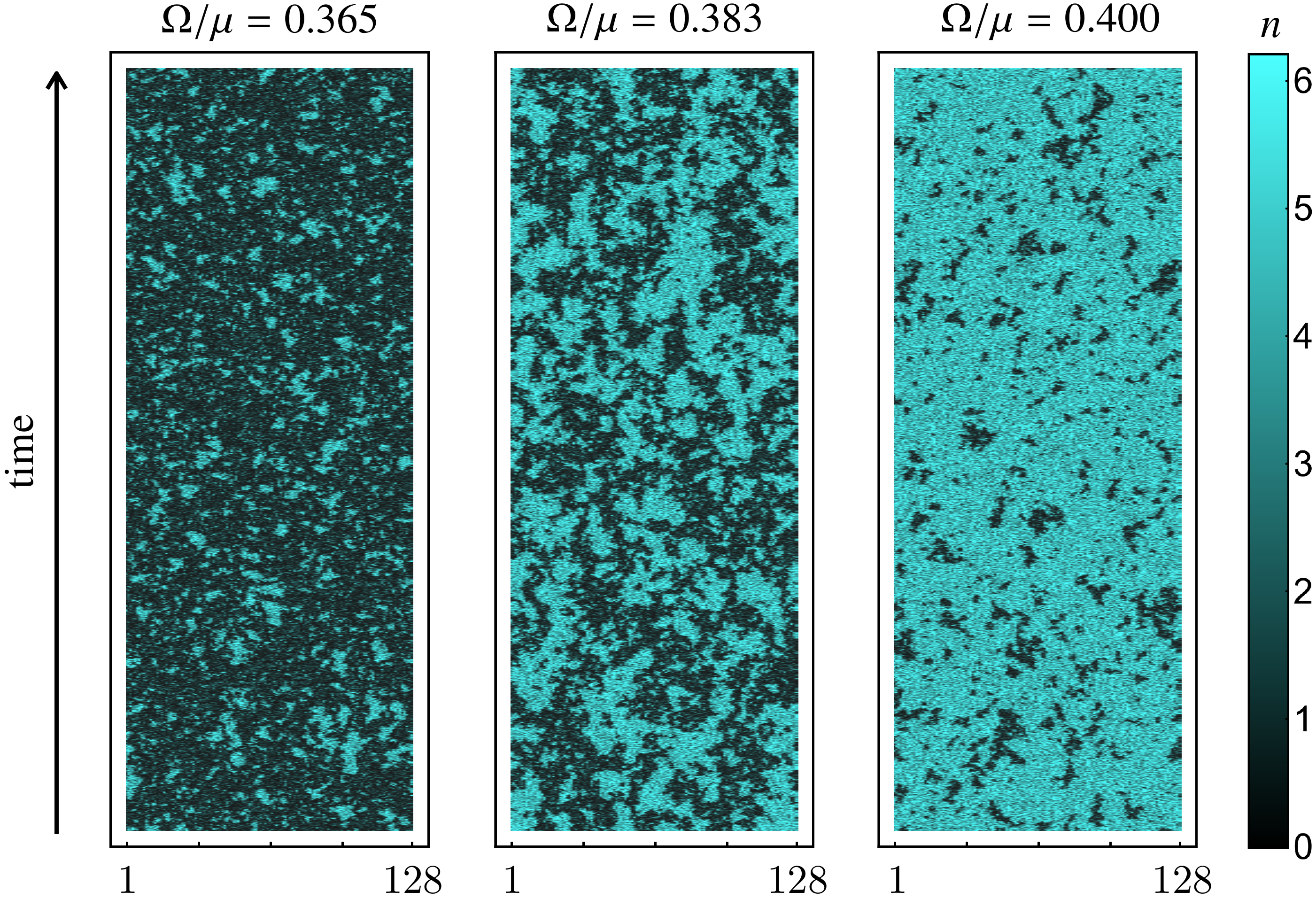}
\caption{(Color online) Real-time dynamics (after burn in) in a 1D system with 128 sites and periodic boundary conditions, showing domain proliferation in the vicinity of the crossover in 1D.  In all three plots, $(J,U,\kappa)\approx(0.1\mu,0.2\mu,0.3\mu)$.  The left panel is just on the dark side of the crossover, the middle panel is roughly in the middle of the crossover, and the right panel is just on the bright side of the crossover.  The color indicates the density.}
\label{fig:dynamics_1d}
\end{figure}

\begin{figure}[!t]
\includegraphics[width=1.0\columnwidth]{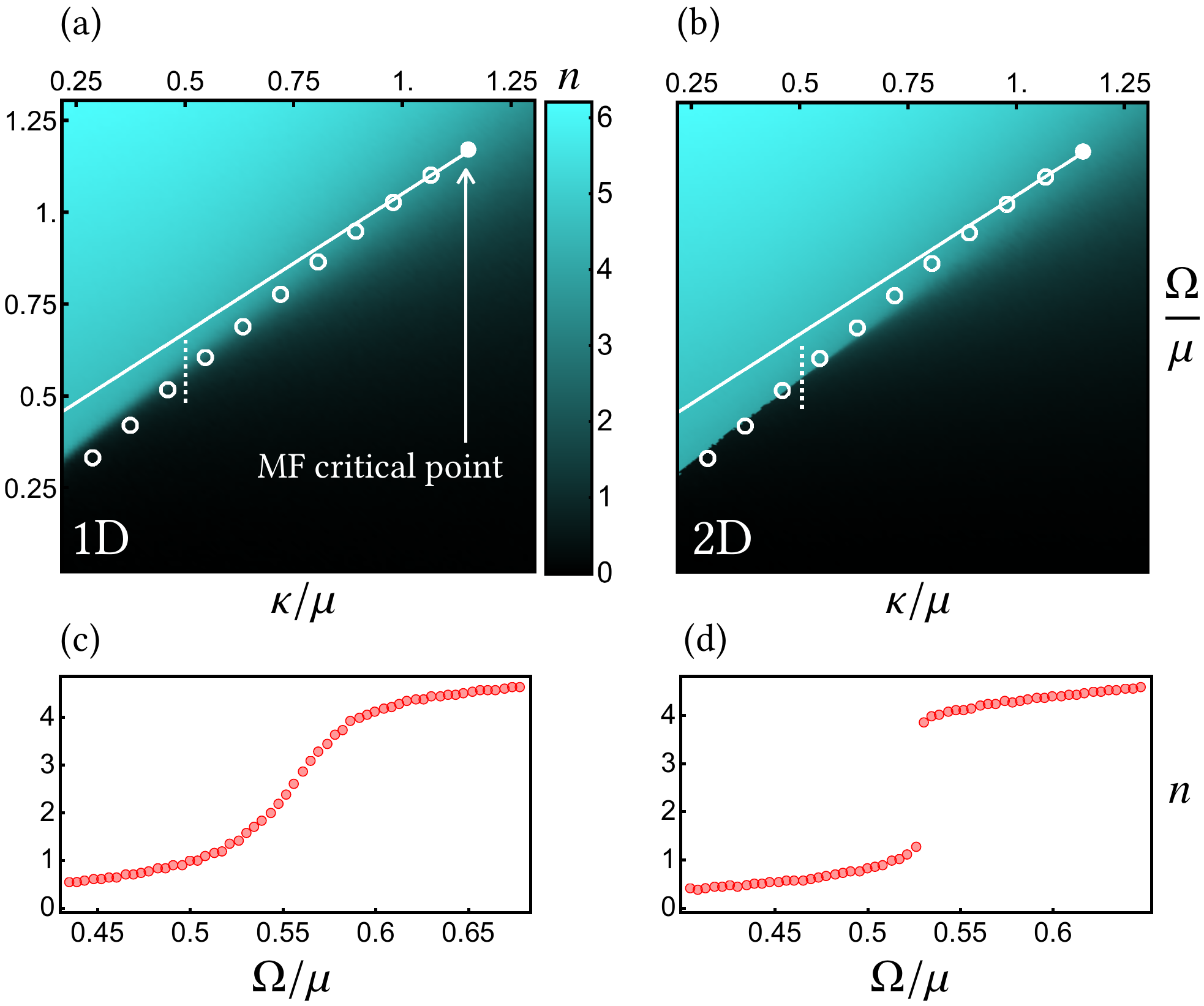}
\caption{(Color online) (a,b) Phase diagrams obtained by numerically solving \eref{eq:sGPE} on (a) a 1D chain with 128 sites and (b) a 2D square lattice with $32\times 32$ sites (in both cases periodic boundary conditions were used).  For both plots, the parameters used are $(J,U)\approx(0.1\mu,0.2\mu)$.  The solid white line locates the (near-critical) condition for a vanishing domain-wall velocity, $h=0$, while the white circles indicate numerically obtained velocity zeros.  The dashed white lines indicate the parameter regime used for plots (c,d).  Note that the phase diagrams are cut off at small $\kappa$, or equivalently low effective temperature, because statistically converged numerical solutions of \eref{eq:sGPE} require prohibitively long integration times as fluctuations become weaker.  (c,d) Cuts through the phase diagrams indicated by dashed white lines in (a) and (b).  In 1D (c) the mean-field phase transition is smoothed out into a crossover, while in 2D (d) bistability leads to a true first-order phase transition. The error bars in (c) and (d) are smaller than the size of the plot markers.}
\label{fig:pd1d}
\end{figure}

Since the above argument relies very little on the dynamics being relaxational, and primarily on the existence of domain walls that---in the absence of fluctuations and for asymptotically large domains---have a directional preference that changes as we move through the bistable region, it is reasonable to expect the qualitative picture described above to be valid even away from the critical point.  Nevertheless, because the dynamics generated by \eref{eq:sGPE} does not induce an equilibrium steady-state distribution far away from the critical point, it is important to verify this picture numerically.  To this end, we carry out a brute-force numerical integration of \eref{eq:vdot_noise} using a fixed time-step first-order Euler-Mayurama method.  After a burn-in time, the equations are integrated until statistical error bars ($1\sigma$) of fractional size 0.01 are achieved for the density.  Temporal autocorrelations on time-scales of $1/10$ the total integration time are also required to fall below a similar threshold to ensure that the averaging time is long compared to all dynamical time-scales, which can become anomalously large near the crossover or phase transition.  Example results of these numerics in one spatial dimension are shown in \fref{fig:dynamics_1d}, and reflect the spatio-temporal dynamics of output light intensity that would be observed if the model were realized in an array of exciton-polariton microcavities.  As expected, by sweeping vertically through the mean-field bistable region, we change from a dominantly dark steady-state with small domains of the bright phase to a predominantly bright steady-state with small domains of the dark phase.  Because the domain walls are unbound, this change manifests as a smooth crossover rather than a true phase transition, as confirmed in the 1D phase diagram shown in \fref{fig:pd1d}a (in particular, see the cross-section plotted in \fref{fig:pd1d}c).  In two dimensions (Figs.\,\ref{fig:pd1d}b and \ref{fig:pd1d}d), there is a clear first-order phase transition between the bright and dark phases, consistent with the expected equilibrium physics of the 2D Ising model.

\section{Discussion\label{sec:conclusion}}

By bringing together a number of ideas from both quantum optics and condensed-matter physics, we have identified a limit of the driven-dissipative Bose-Hubbard model in which the dominant fluctuations are captured by non-equilibrium Langevin equations, enabling a quantitatively accurate and computationally efficient determination of steady-state properties.  Near the critical point, these fluctuations are thermal and lead to an effective equilibrium description.  However, we emphasize that the Langevin description of the problem should be asymptotically exact in the limit of $\mathcal{N}\rightarrow\infty$ even away from the critical point, where an equilibrium description is \emph{not} valid.  Numerically, we find that in 2D the first-order phase transition expected from the mapping onto an Ising model remains in tact far from the critical point, where this mapping is not strictly valid.  In addition, the absence of a phase transition in 1D is the result of the same domain-wall phenomenology that prevents ordering of the 1D Ising model at finite temperatures.  In this way, the numerical results reinforce and extend the assertion that the steady-state behavior of the driven-dissipative Bose-Hubbard model possesses an emergent description in terms of the equilibrium physics of a finite-temperature classical Ising model.  These conclusions have direct consequences for a range of experiments in which particle loss is countered by coherent driving, for example a coherently driven exciton-polariton fluid in an array of semiconductor microcavities.  Here, by tuning the laser driving strength through the mean-field bistable regime, one should be able to observe domain growth, hysteresis, critical fluctuations, and other generic features of the Ising model in a longitudinal field, all by simple correlation measurements on the output intensities of the cavities.

We caution that any claims about the universality class of the phase transition require more than just a microscopically accurate treatment of fluctuations---it is also important to identify the relevance of any ignored fluctuations in the sense of the renormalization group, even if they are parametrically small.  In other words, there is no guarantee that small ($1/\mathcal{N}$) quantitative errors at the scale of the lattice spacing will not qualitatively affect the nature of the phase transition.  Renormalization group arguments in favor of the Ising universality class have already been presented \cite{PhysRevB.93.014307}, however the status of the first-order phase transition away from the critical point is less clear.  It would be worthwhile to extend the numerical approach taken in \sref{sec:langevin} to confirm the universal aspects of both the critical point and the first-order phase transition.

It would also be worthwhile to compare some of the results in 1D with numerically exact calculations based on the density-matrix renormalization group  \cite{doi_10_1080_00018732_2014_933502} in order to better understand the importance of higher-order (in $1/\mathcal{N}$) corrections that are not captured by the Langevin description.  In particular, at large $U$ (small $\mathcal{N}$), mean-field arguments suggest that the inclusion of fluctuations ignored to leading order in $1/\mathcal{N}$ may lead to richer steady-state behaviors \cite{PhysRevLett.110.233601,PhysRevA.90.063821}, including phases that spontaneously break discrete spatial-translation symmetry \cite{PhysRevLett.110.233601,PhysRevA.94.033801}.   In 2D, the formalism described here could be used to compute other dynamical aspects of the system near the first-order phase transition; for example, it should be possible to calculate the lifetime of the metastable phase via an instanton approach.

\section{Acknowledgments}

We thank Cristiano Ciuti, Sebastian Diehl, Howard Carmichael, Sarang Gopalakrishnan, Victor Gurarie, Ana Maria Rey, Murray Holland, Anzi Hu, Michael Fleischhauer, and Chih-Wei Lai for helpful discussions.  M.F.M., J.T.Y., and A.V.G. acknowledge support by ARL CDQI, ARO MURI, NSF QIS, ARO, NSF PFC at JQI, and AFOSR. R.M.W. acknowledges partial support from the National Science Foundation under Grant No. PHYS- 1516421.  M.H. acknowledges support by AFOSR-MURI, ONR and Sloan Foundation.

\appendix

\section{Path integral formalism\label{sec:appendix_A}}

The path integral presented in \sref{sec:formalism} is closely related to the usual Keldysh functional integral formalism, for which there are many good references (see for example Ref.\,\cite{book_kamenev}).  However, there is a subtle difference between the formalism used here and that typically employed in the condensed matter community, and it therefore seems worthwhile to provide an explicit derivation of Eqs.\,(\ref{eq:partition_function}-\ref{eq:psi_eq_time}).  For simplicity, we will treat only the single-cavity case, but the generalization to many cavities that yields Eqs.\,(\ref{eq:partition_function}-\ref{eq:psi_eq_time}) follows immediately.

In any path integral formulation of quantum mechanics, operators must be traded in for classical variables.  The usual way to do this, as is the case in the standard approach to the Keldysh function integral, is to repeatedly insert coherent-state resolutions of identity during the time-evolution.  Operators get sandwiched between coherent states, and if they are normal-ordered they turn into functions of phase space variables.  In the language of quantum optics, operators are exchanged for their $Q$-symbols.  However, there are many different ways to associate operators with functions over phase space, and thus many ways to formulate a path integral.  In the following derivation, we will replace operators with classical variables by working in the Weyl representation (see for example Ref.\,\cite{PhysRevA.80.033624}).  In our opinion, even though this strategy entails some additional overhead in phase-space formalism, it is both more direct and conceptually simpler than the usual approach to the Keldysh functional integral.  In particular, the canonical approach described in Refs.\,\cite{doi:10.1080/00018730902850504,book_kamenev} relies on the construction of a formal continuous-time notation that---together with simple rules for computing equal-time correlation functions---correctly reproduces the continuous-time limit of the Greens functions of a non-interacting Bose field. Interactions are then included in a self-consistent fashion by ensuring that the rules for Gaussian integration produce correct results for the interacting theory at all order of perturbation theory.  In the approach taken here, the path integral is derived constructively in such a way that an unambiguous continuous-time notation emerges naturally from a properly defined (i.e. discretized) functional integral.

\subsection{Representation of the Wigner function}
The Weyl-symbol of an arbitrary operator $\hat{A}$, denoted $\mathcal{A}_{\rm w}(\psi)$, can be defined via the relation
\begin{align}
\mathcal{A}_{\rm w}(\psi)={\rm Tr}\big(\delta_{\rm w}(\psi-\hat{a})\hat{A}\big).
\end{align}
Here, the Weyl-ordered (and operator valued) delta function is defined by
\begin{align}
\delta_{\rm w}(\psi-\hat{a})=\frac{1}{\pi^2}\int d^2\varphi\exp\Big(\bar{\varphi}\big(\psi-\hat{a}\big)-\varphi\big(\bar{\psi}-\hat{a}^{\dagger}\big)\Big).
\end{align}
When convenient, the correspondence between operators and their Weyl symbols will be indicated below with the notation $\hat{A}\leftrightarrow\mathcal{A}_{\rm w}(\psi)$.  Given the special role played by the density operator $\hat{\rho}$, it is traditional to use a special notation for its Weyl symbol, $\mathcal{W}(\psi,t)$, which is also called the Wigner function; the explicit time-dependence is included because we will work in the Schr\"odinger picture, where the density matrix (and therefore the Wigner function) evolves in time.

The Weyl representation is intimately related to symmetrically ordered operator products; if an arbitrary operator $\hat{A}$ is expanded in terms of symmetrically-ordered operator products,
\begin{align}
\label{eq:O_pq}
\hat{A}=\sum_{p,q}\mathcal{A}_{pq}[\hat{a}^p(\hat{a}^{\dagger})^q]_{\rm s},
\end{align}
then the coefficients in this expansion determine the Weyl symbol $\mathcal{A}_{\rm w}(\psi)$ in a particularly natural way,
\begin{align}
\label{eq:wigner_expansion}
\mathcal{A}_{\rm w}(\psi)=\sum_{p,q}\mathcal{A}_{pq}\psi^p\bar{\psi}^q.
\end{align}

Given the Wigner function at an initial time $t_0$, we would like to understand how it has changed a short time $\delta t$ later due to the evolution of the density matrix by the master equation.  During this time, the density matrix evolves according to $\hat{\rho}(t_0+\delta t)=V_{\delta t}(\hat{\rho}(t_0))$, where the infinitesimal time evolution (super-)operator $V_{\delta t}$ satisfies
\begin{align}
\label{eq:meq_inf}
V_{\delta t}(\star)&=1-i\delta t[\hat{H},\star]\\
&+\delta t\frac{\kappa}{2}\sum_{j}\big(2\hat{a}^{\phantom\dagger}_j\star \hat{a}^{\dagger}_j-\star\hat{a}^{\dagger}_j\hat{a}^{\phantom\dagger}_j-\hat{a}^{\dagger}_j\hat{a}^{\phantom\dagger}_j\star\big)+\mathcal{O}(\delta t^2).\nonumber
\end{align}
This transformation induces a corresponding evolution of the Wigner function, which for now we write formally as
\begin{align}
\label{eq:weq_inf}
\mc{W}(\psi_1,t_0+\delta t)=\int d^2\psi_0\,\mathcal{V}(\psi_1,\psi_0)\mc{W}(\psi_0,t_0),
\end{align}
thereby implicitly defining the infinitesimal phase-space propagator for the Wigner function, $\mathcal{V}$.

From the structure of Eqs.\,(\ref{eq:meq_inf},\ref{eq:weq_inf}), it is clear that finding the explicit form of $\mathcal{V}$ requires us to compute the Weyl symbol of products of $\hat{\rho}$ with creation and annihilation operators.  To this end we define an operator-valued generating function
\begin{align}
\hat{G}=e^{\eta\hat{a}+\bar{\eta}\hat{a}^{\dagger}}\hat{\rho}(t_0),
\end{align}
which can be differentiated to produce symmetrically ordered operator products
\begin{align}
\label{eq:op_ps_correspondence}
\partial^p_{\eta}\partial^q_{\bar{\eta}}\hat{G}\big|_{\eta=0}=[\hat{a}^p(\hat{a}^{\dagger})^q]_{\rm s}\hat{\rho}(t_0).
\end{align}
Using \eref{eq:O_pq} and \eref{eq:op_ps_correspondence}, we can expand the product of an arbitrary operator with the density matrix as
\begin{align}
\hat{A}\hat{\rho}(t_0)=\sum_{p,q}\mathcal{A}_{pq}\partial^p_{\eta}\partial^q_{\bar{\eta}}\hat{G}\big|_{\eta=0},
\end{align}
giving us a prescription to compute the Weyl symbol of $\hat{A}\hat{\rho}(t_0)$ from the Weyl symbol of $\hat{G}$
\begin{align}
\label{eq:mult_corr}
\hat{A}\hat{\rho}(t_0)\leftrightarrow \sum_{p,q}\mathcal{A}_{pq}\partial^p_{\eta}\partial^q_{\bar{\eta}}\mathcal{G}_{\rm w}(\psi_1)\big|_{\eta=0}.
\end{align}
Making use of the standard operator phase-space correspondences \cite{book_carmichael_1},
\begin{align}
&\hat{a}\hat{\rho}\leftrightarrow(\psi+\frac{1}{2}\partial_{\bar{\psi}})\mathcal{W}(\psi),~~~~~\hat{a}^{\dagger}\hat{\rho}\leftrightarrow(\bar{\psi}-\frac{1}{2}\partial_{\psi})\mathcal{W}(\psi),\nonumber\\
&\hat{\rho}\hat{a}\leftrightarrow(\psi-\frac{1}{2}\partial_{\bar{\psi}})\mathcal{W}(\psi),~~~~~\hat{\rho}\hat{a}^{\dagger}\leftrightarrow(\bar{\psi}+\frac{1}{2}\partial_{\psi})\mathcal{W}(\psi),\nonumber
\end{align}
the commutation relations $[\partial_{\psi},\psi]=1$, $[\partial_{\psi},\bar{\psi}]=0$, and the Baker-Campbell-Hausdorff formula,
the Weyl symbol of $\hat{G}$ can be written
\begin{align}
\label{eq:G_is_1}
\mathcal{G}_{\rm w}(\psi_1)=e^{\frac{1}{2}\eta\partial_{\bar{\psi}_1}-\frac{1}{2}\bar{\eta}\partial_{\psi_1}}e^{\eta\psi_1+\bar{\eta}\bar{\psi}_1}\mathcal{W}(\psi_1,t_0).
\end{align}
Inserting a standard representation of the delta function and integrating by parts, we obtain
\begin{align}
\label{eq:G_is_2}
&\mathcal{G}_{\rm w}(\psi_1)=\frac{4}{\pi^2}\int d^2\varphi_0 d^2\psi_0\times\\
&e^{2\varphi_0(\bar{\psi}_1-\bar{\psi}_0)-2\bar{\varphi}_0(\psi_1-\psi_0)}e^{\eta(\varphi_0+\psi_0)+\bar{\eta}(\bar{\varphi}_0+\bar{\psi}_0)}\mathcal{W}(\psi_0,t_0)\nonumber.
\end{align}
Note that the choice of a generating function that produced symmetrically ordered operator products also led to all of the derivatives appearing on the left in the first line of \eref{eq:G_is_1}, which enabled the integration by parts to proceed in a particularly simple manner to obtain \eref{eq:G_is_2}.  Inserting \eref{eq:G_is_2} into \eref{eq:mult_corr} and then using \eref{eq:wigner_expansion}, we obtain
\begin{align}
\label{eq:wigner_prod}
&\hat{A}\hat{\rho}\leftrightarrow\\
&\frac{4}{\pi^2}\int d^2\varphi_0 d^2\psi_0e^{2\varphi_0(\bar{\psi}_1-\bar{\psi}_0)-2\bar{\varphi}_0(\psi_1-\psi_0)}\mathcal{A}_{\rm w}(\psi_0+\varphi_0)\mathcal{W}(\psi_0,t_0).\nonumber
\end{align}

Given \eref{eq:wigner_prod}, we can now deduce \eref{eq:weq_inf} from \eref{eq:meq_inf}.  To first order in $\delta t$ we find
\begin{align}
\label{eq:W_evolve_inf}
\mathcal{W}(\psi_1,t_0+\delta t)=\frac{4}{\pi^2}\int d^2\psi_0 d^2\varphi_0 e^{i\delta t\mathcal{L}(\psi_1,\varphi_1;\psi_0,\varphi_0)}\mathcal{W}(\psi_0,t_0),
\end{align}
where
\begin{align}
\mathcal{L}(\psi_1,\varphi_1;\psi_0,\varphi_0)&=2i\bar{\varphi}_0(\psi_1-\psi_0)/\delta t-2i\varphi_0(\bar{\psi}_1-\bar{\psi}_0)/\delta t\nonumber\\
&-\mathcal{H}_{\rm w}(\psi_0+\varphi_0)+\mathcal{H}_{\rm w}(\psi_0-\varphi_0)\nonumber\\
&+i\kappa(2\bar{\varphi}_0\varphi_0-\varphi_0\bar{\psi}_0+\bar{\varphi}_0\psi_0).
\end{align}
The Wigner function at a general time $t$ can be obtained from the Wigner function at time $t_0$ by iteration of \eref{eq:W_evolve_inf}.  Breaking the interval $[t_0,t]$ into $N$ segments of size $\delta t=(t-t_0)/N$, we obtain
\begin{align}
\mathcal{W}(\psi_N,t)=\int\prod_{j=0}^{N-1}\bigg(\frac{4}{\pi^2}d^2\psi_{j}d^2\varphi_{j}\bigg)e^{i\mathcal{S}}\mathcal{W}(\psi_0,t_0).
\end{align}
Here we have defined the discretized action
\begin{align}
\label{eq:discrete_action}
\mathcal{S}=\sum_{j=0}^{N-1}\delta t\mathcal{L}(\psi_{j+1},\varphi_{j+1};\psi_j,\varphi_j).
\end{align}
Defining functional integration measures that include the fields $\psi_{N},\varphi_N$ at the final time,
\begin{align}
\label{eq:discrete_functional_measure}
\mathscr{D}\psi=\prod_{j=0}^{N}d^2\psi_{j},~~~~\mathscr{D}\varphi=\prod_{j=0}^{N}\frac{4}{\pi^2}d^2\varphi_{j},
\end{align}
the trace of the Wigner function at time $t$ can now be written
\begin{align}
\label{eq:discrete_partition_function}
\mathcal{Z}\equiv\int \mathscr{D}\psi\mathscr{D}\varphi e^{i \mathcal{S}}\mathcal{W}(\psi_0,t_0)=1.
\end{align}
The continuous-time limit ($N\rightarrow\infty,\delta t\rightarrow0$) of Eqs.\,(\ref{eq:discrete_action}-\ref{eq:discrete_partition_function}), generalized to many coherently coupled bosonic modes, yields Eqs.\,(\ref{eq:partition_function},\ref{eq:action}) of the main text.

\subsection{Expectation values}

\begin{figure}[!h]
\includegraphics[width=1.0\columnwidth]{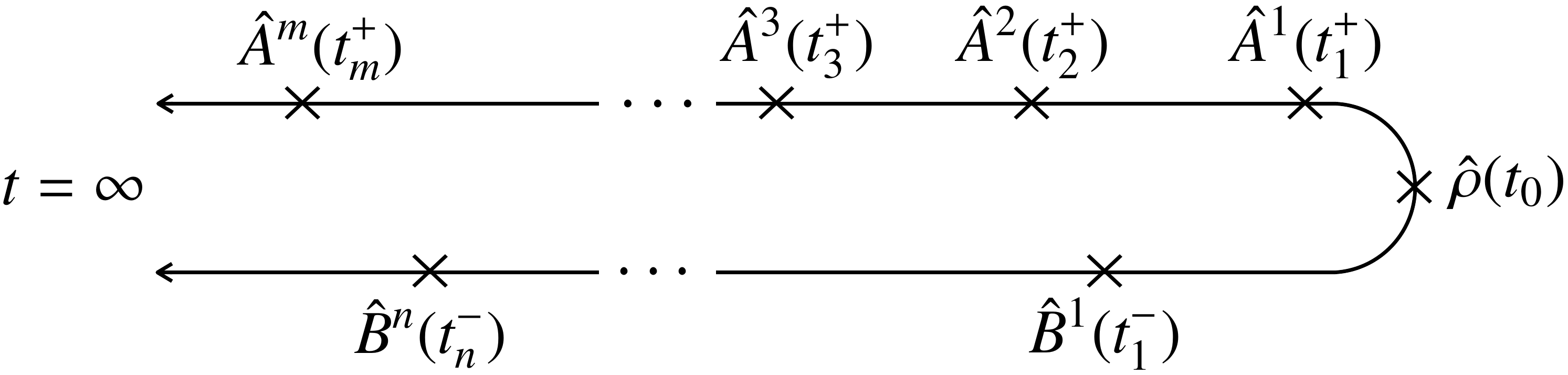}
\caption{Time ordering of operators in \eref{eq:keldysh_order}.}
\label{fig:keldysh}
\end{figure}

The functional integral representation of the Wigner function lends itself naturally to calculating correlation functions that are time-ordered along the Keldysh contour (\fref{fig:keldysh}), e.g.
\begin{align}
\label{eq:keldysh_order}
\mathscr{C}=\big\langle\mathcal{T}_K\big(\hat{B}^1(t^-_1)\dots\hat{B}^{n}(t^-_n)\hat{A}^{1}(t^+_1)\dots\hat{A}^{m}(t^+_m)\big)\big\rangle.
\end{align}

Here, $\mathscr{C}$ is written in the Heisenberg picture, and $\mathcal{T}_{K}$ orders operators at times with a ``$+$'' superscript such that the times increase from right to left, and orders operators at times with a ``$-$'' superscript such that the times increase from left to right.  Note that the Heisenberg picture is to be interpreted \emph{before} the adiabatic elimination of the reservoir, which we assume proceeds within the Born-Markov approximation.  Assuming without loss of generality that $t^{\pm}_j>t^{\pm}_{j-1}$, insisting that the operators be inserted at the discretized time-slices chosen above, ($t^{\pm}_j= \delta t\times r^{\pm}_j$, with $r_j^{\pm}$ an integer between $0$ and $N$), and making the notational change $\hat{A}(\delta t r^{\pm}_j)\rightarrow\hat{A}(r^{\pm}_j)$, $\mathscr{C}$ can be written
\begin{align}
\mathscr{C}={\rm Tr}\Big(\hat{A}^m(r^+_m)\dots\hat{A}^1(r^+_1)\,\hat{\rho}(t_0)\,\hat{B}^1(r^-_1)\dots\hat{B}^n(r^-_n)\Big).
\end{align}
For our purposes we need to rewrite this correlation function in the Schr\"odinger picture, which can be accomplished with the help of the quantum regression formula.  For example, if $t_1^+<t_1^-$, we have \cite{book_carmichael_1}
\begin{align}
\mathscr{C}={\rm Tr}\Big(\dots U_{\delta t(r_{1}^{-}-r_{1}^{+})}\big(\hat{A}^1U_{\delta t r_1^+}\big(\hat{\rho}(t_0)\big)\big)\hat{B}^1\dots\Big).
\end{align}
The choice of Keldysh-ordering is necessary and sufficient to guarantee that, in the quantum regression formula, it will never be necessary to evolve backwards in time. Using the path-integral expression for the Wigner function at time $t$, together with the phase space correspondences for operators $\hat{A}$ and $\hat{B}$, we find
\begin{align}
\mathscr{C}=\int &\mathscr{D}\psi\mathscr{D}\varphi \,e^{-i\mathcal{S}}\mc{W}(\psi_0,t_0)\times\\
&\big(\dots\mathcal{A}_{\rm w}^1(\psi_{r_1^+}+\varphi_{r_1^+})\mathcal{B}_{\rm w}^1(\psi_{r_1^-}-\varphi_{r_1^-})\dots\big)\\
&\equiv\big\langle\dots\mathcal{A}_{\rm w}^1(\psi_{r_1^+}+\varphi_{r_1^+})\mathcal{B}_{\rm w}^1(\psi_{r_1^-}-\varphi_{r_1^-})\dots\big\rangle_{\mc{Z}}.
\end{align}
Such correlation functions can be conveniently computed with respect to a generating functional by adding source terms to action,\begin{align}
\mathscr{C}=&\bigg(\dots\mathcal{A}_{\rm w}^1\Big(\frac{\partial}{\delta t\partial J_{r_1^+}}+\frac{\partial}{\delta t\partial K_{r_1^+}}\Big)\mathcal{B}_{\rm w}^1\Big(\frac{\partial}{\delta t\partial J_{r_1^-}}-\frac{\partial}{\delta t\partial K_{r_1^-}}\Big)\dots\bigg)\nonumber\\
&\times\mathcal{Z}(\bm{J},\bm{K})\Big|_{\bm{J},\bm{K}=0}\nonumber.
\end{align}
Here
\begin{align}
\mathcal{Z}(\bm{J},\bm{K})=\int &\mathscr{D}\psi\mathscr{D}\varphi \,e^{i\mathcal{S}(\bm{J},\bm{K})}\mc{W}(\psi_0,t_0),
\end{align}
and
\begin{align}
\mathcal{S}(\bm{J},\bm{K})=\mathcal{S}+\sum_{j=0}^{N}\delta t\big(\bm{J}_{j}\cdot\bm{\psi}_{j}+\bm{K}_{j}\cdot\bm{\varphi}_{j}\big),
\end{align}
with
\begin{align}
\bm{J}_j\!=\!\Bigg\lgroup\begin{matrix} J_j \vspace{0.1 cm} \\ \bar{J}_j \end{matrix}\Bigg\rgroup\!,~
\bm{K}_j\!=\!\Bigg\lgroup\begin{matrix} K_j \vspace{0.1 cm} \\ \bar{K}_j \end{matrix}\Bigg\rgroup,~
\bm{\psi}_j\!=\!\Bigg\lgroup\begin{matrix} \psi_j \vspace{0.1 cm} \\ \bar{\psi}_j \end{matrix}\Bigg\rgroup\!,~
\bm{\varphi}_j\!=\!\Bigg\lgroup\begin{matrix} \varphi_j \vspace{0.1 cm} \\ \bar{\varphi}_j \end{matrix}\Bigg\rgroup.
\end{align}
Restricting to the special case when the operators $\hat{A}$ and $\hat{B}$ are creation and annihilation operators, defining
\begin{align}
\frac{1}{\delta t}\frac{\partial}{\partial{J_j}}=\frac{\delta}{\delta J(t_j)},~~~~\frac{1}{\delta t}\frac{\partial}{\partial{K_j}}=\frac{\delta}{\delta K(t_j)},
\end{align}
and taking the continuum limit $\delta t\rightarrow 0$, we recover the expressions for correlation functions in \sref{sec:formalism} of the main text.

The correspondence given above can also be reversed in such a way to rewrite expectation values of the fields $\psi$ and $\varphi$ in terms of operator averages.  To this end, we rearrange the operator phase-space correspondences as
\begin{align}
&\{\hat{a},\hat{\rho}\}\leftrightarrow2\psi\mathcal{W},~~~~~\{\hat{a}^{\dagger},\hat{\rho}\}\leftrightarrow2\bar{\psi}\mathcal{W}\nonumber\\
&[\hat{a},\hat{\rho}]\leftrightarrow\partial_{\bar{\psi}}\mathcal{W},~~~~~[\hat{a}^{\dagger},\hat{\rho}]\leftrightarrow-\partial_{\psi}\mathcal{W}\nonumber
\end{align}
The substitutions $\partial_{\psi}\leftrightarrow-2\bar{\varphi}$ and $\partial_{\bar{\psi}}\leftrightarrow2\varphi$ are valid under the functional integration sign, so long as the derivative sits to the left of any other instances of the field $\psi$ evaluated at the same time (not including those occurring in the action).  Thus we are led to the identifications
\begin{align}
\label{eq:ps_cor_alt}
&\psi\mathcal{W}\leftrightarrow\frac{1}{2}\{\hat{a},\hat{\rho}\},~~~~~\varphi\mathcal{W}\leftrightarrow\frac{1}{2}[\hat{a},\hat{\rho}],
\end{align}
with the understanding that when products of the fields $\psi$ and $\varphi$ at the same time arise, we should take the commutators \emph{after} the anticommutators in the corresponding operator expectation values.  This identification leads very directly to correlation functions of the classical field.  For example, assuming $r_3>r_2>r_1$ we have
\begin{align}
\label{eq:equal_time}
\big\langle \bar{\psi}_{r_1}\psi_{r_2}\bar{\psi}_{r_3}\big\rangle_{\mc{Z}}={\rm Tr}\big(\{\hat{a}^{\dagger}(r_3),\{\hat{a}(r_2),\{\hat{a}^{\dagger}(r_1),\hat\rho(t_0)\}\}\}\big).
\end{align} 
Note that if $r_1=r_2=r_3\equiv r$, this correlation function simplifies to
\begin{align}
\label{eq:psi_eq_time_sup}
\big\langle \bar{\psi}_{r}\psi_{r}\bar{\psi}_{r}\big\rangle_{\mc{Z}}={\rm Tr}\big([\hat{a}^{\dagger}(r)\hat{a}(r)\hat{a}^{\dagger}(r)]_s\big).
\end{align} 
Since we just have a few fields in the correlation function, this result can be worked out by direct comparison of the right-hand sides of \eref{eq:equal_time} and \eref{eq:psi_eq_time_sup} (using the commutation relation $[\hat{a},\hat{a}^{\dagger}]=1$).  More generally, it also follows by using the phase-space correspondence $[\hat{a}^{\dagger}\hat{a}\hat{a}^{\dagger}]_s\leftrightarrow (\bar\psi+\bar\varphi)^2(\psi+\varphi)$.  Going back to operator expectation values by using \eref{eq:ps_cor_alt}, and remembering the rule that commutators are taken after anticommutators when the fields $\varphi$ and $\psi$ are evaluated at the same times, terms with one or more power of $\varphi$ result in an outer commutator that gets killed by the trace, so only the $\bar{\psi}^2\psi$ term survives.  The generalization of \eref{eq:psi_eq_time_sup} to arbitrary equal-time correlation functions of the classical field $\psi$ leads, in the continuum-time limit and for more than one site, to \eref{eq:psi_eq_time} of the main text.

\section{Adiabatic elimination of the massive field \label{sec:appendix_B}}

Here we describe the perturbative adiabatic elimination of the field $\varrho$ near the mean-field critical point.  For simplicity, we will first treat the case with no fluctuations ($\mc{N}\rightarrow\infty$), and afterwards we will consider the effect of weak fluctuations.
\subsection{Mean-field theory}
Substitution of \eref{eq:psi_decomp} into \eref{eq:mft_lattice} yields coupled equations for $\varrho$ and $\sigma$ that are fully equivalent to the mean-field dynamics of $\Psi$.  To simplify the following expressions we convert all energy/time scales into dimensionless rations with the chemical potential $\mu$, and in a slight abuse of notation we do not change any of the associated symbols---the $\mu$-dependence can be unambiguously restored by insisting on dimensional consistency.  After some algebra, we find
\begin{align}
\dot\varrho&=-\frac{2}{\sqrt{3}}(\varrho-\varrho_0)-r\varrho-\frac{J\nabla^2}{\sqrt{3}}\big(\varrho+2\sigma\big)\nonumber\\
\label{eq:rho_expand}
&+\frac{u}{\sqrt{3}}\big(2\sigma^3+3\sigma\varrho^2+3\varrho\sigma^2+\varrho^3-\sigma^2\sqrt{6/u}\big),\\
\dot{\sigma}&=-\frac{h}{2}-r\sigma +\frac{J\nabla^2}{\sqrt{3}}\big(\sigma+2\varrho\big)\nonumber\\
\label{eq:sigma_expand}
&-\frac{u}{\sqrt{3}}\big(\varrho^2\sqrt{6/u}+2\varrho^3+3\sigma\varrho^2+3\varrho\sigma^2+\sigma^3\big),
\end{align}
where
\begin{align}
\varrho_0=\frac{h\sqrt{3}}{8}-\frac{r}{\sqrt{8u}}.
\end{align}
Several simplifications can now be made.  First, because we are interested in dynamics near the critical point and after the field has nearly relaxed, $h$, $r$, $\varrho$, and $\sigma$ can all be treated as small parameters.  Though we do not know \emph{a priori} how small the fields $\varrho$ and $\sigma$ are relative to the parameters $r$ and $h$, it is perfectly consistent to keep, at any particular order in one of the parameters, only the lowest non-trivial order in any other parameter.  Moreover, while we don't know how small spatial derivatives of the field are, we do expect them to be small compared to the fields themselves, which should vary slowly in space near the critical point, and thus we formally treat $J\nabla^2$ as an additional small parameter (one can check that this assumption is self consistent at the end of the calculation, where it will be seen that $J\nabla^2\sim r$).  Following this logic, and introducing the rescaled parameters $K=J/\sqrt{3}$ and $g=u/\sqrt{3}$ used in the main text, we arrive at the simplified equations
\begin{align}
\label{eq:eqm_rho}
\dot{\varrho}&=-\frac{2}{\sqrt{3}}(\varrho-\varrho_0)-2K\nabla^2\sigma+\mathcal{O}(\sigma^2),\\
\label{eq:eqm_sigma}
\dot{\sigma}&=-\frac{h}{2}-r\sigma+K\nabla^2\big(\sigma+2\varrho\big)-g\sigma^3+\mathcal{O}(\varrho^2)+\mathcal{O}(\varrho \sigma^2).
\end{align}
Terms that are kept with the $\mathcal{O}$ notation are there to remind us that we do not know for sure whether they are parametrically small compared to other terms that are kept---they will turn out to be unimportant for reasons explained below, which is why we don't keep track of the exact coefficients.

The justification for adiabatically eliminating $\varrho$ near the critical point is now clear: As $r\rightarrow 0$, the term proportional to $\varrho$ in \eref{eq:eqm_rho} stays finite, indicating that $\varrho$ relaxes to zero exponentially in time even at the critical point (once $\mu$ is restored, we see that it decays on a timescale $\sim 1/\mu$).  On the other hand, the term linear in $\sigma$ in \eref{eq:eqm_sigma} vanishes as $r\rightarrow 0$, indicating a divergent timescale for relaxation of $\sigma$ (which relaxes algebraically precisely at the critical point, $r=0$).  To adiabatically eliminate $\varrho$ we set the r.h.s of \eref{eq:eqm_rho} to zero, obtaining
\begin{align}
\varrho=\varrho_0-\sqrt{3}K\nabla^2\sigma+\mathcal{O}(\sigma^2),
\end{align}
and then substitute this result into \eref{eq:eqm_sigma}.  Many derivative terms are generated, but working to lowest order in $K\nabla^2$ we find
\begin{align}
\label{eq:sigma_simp}
\dot\sigma&=K\nabla^2\sigma-r\sigma-g\sigma^3-\frac{h}{2}+\mathcal{O}(r^2)+\mathcal{O}(r \sigma^2).
\end{align}
Here we have implicitly assumed that $|h|<|r|$ to write $\varrho_0=\mathcal{O}(r)$, which poses no important restriction on what follows.  If we ignore the final two terms and solve \eref{eq:sigma_simp} at $h=0$ and $r<0$ (inside the bistable region), we find two uniform solutions at $\sigma=\pm\sqrt{|r|/g}$, which sets the scale of $\sigma$ in the relevant near-critical dynamics, $\sigma\sim\sqrt{r}$.  From this scaling, it is easily seen that the final two terms in \eref{eq:sigma_simp} are parametrically smaller than the others; thus we were justified in dropping them, which results in \eref{eq:vdot} in the main text.

\subsection{Fluctuations}

Substitution of \eref{eq:psi_decomp} into \eref{eq:sGPE} yields coupled stochastic equations for $\varrho$ and $\sigma$ that are fully equivalent to the non-equilibrium Langevin equation for $\Psi$.  In the limit of weak noise, an expansion in $\varrho$ and $\sigma$ is still justified.  Moreover, since arbitrarily weak noise will induce arbitrarily small excursions away from the mean-field stationary state, much of the analysis that lead to an effective relaxational description of $\sigma$ (that is approximately decoupled from $\varrho$) should remain valid, as it assumed nothing more than being close to both the critical point and the steady state.  Working near the mean-field critical point, the same assumptions that lead from Eqs.\,(\ref{eq:rho_expand}-\ref{eq:sigma_expand}) to Eqs.\,(\ref{eq:eqm_rho}-\ref{eq:eqm_sigma}) remain justified and yield (for now keeping all gradient terms)
\begin{align}
\label{eq:cant_eliminate_rho}
\dot\varrho&=-\frac{2}{\sqrt{3}}(\varrho-\varrho_0)-K\nabla^2(\varrho+2\sigma)+\mathcal{O}(\sigma^2)+\eta(\tau),\\
\label{eq:need_to_simplify}
\dot\sigma&=-\frac{h}{2}-r\sigma+K\nabla^2(\sigma+2\varrho)-g\sigma^3+\mathcal{O}(\varrho^2)+\mathcal{O}(\varrho \sigma^2)+\xi(\tau).
\end{align}
In order to avoid confusion regarding the noise variances, here we have chosen a new symbol for the dimensionless time, $t\mu\equiv\tau$ (so $\dot\sigma= d\sigma/d\tau$, etc.), even though we continue to use the same symbols for the (now dimensionless) energies $r$, $h$, $K$, and $g$.  The real noises $\eta$ and $\xi$ are defined
\begin{align}
\eta(\tau)=\frac{\zeta_{\rm I}(t)}{\mu}-\frac{\zeta_{\rm R}(t)}{\mu\sqrt{3}},~~~\xi(\tau)=\frac{2\zeta_{\rm I}(t)}{\mu\sqrt{3}},
\end{align}
where $\zeta_{\rm R}$ and $\zeta_{\rm I}$ are the real and imaginary components, respectively, of the complex Gaussian white noise $\zeta(t)$ in \eref{eq:sGPE}. Thus from the variances of $\zeta(t)$ we have
\begin{align}
\overline{\eta_j(\tau_1)\eta_k(\tau_2)}=\overline{\xi_j(\tau_1)\xi_k(\tau_2)}=\frac{\kappa}{3\mathcal{N}}\delta_{j,k}\delta(\tau_1-\tau_2),
\end{align}
where one factor of $1/\mu$ has been absorbed into the now dimensionless $\kappa$ and one is used to change variables from $t$ to $\tau$ in the $\delta$ function.

We would like to eliminate the terms in \eref{eq:need_to_simplify} that involve $\varrho$, but, strictly speaking, the adiabatic elimination of $\varrho$ by setting $\dot\varrho=0$ in \eref{eq:cant_eliminate_rho} is no longer justified.  Nevertheless, for weak noise and near the critical point, there will be a separation of timescales, length-scales, and typical sizes of the fluctuations of $\sigma$ and $\varrho$. Because $\varrho$ remains massive at the mean-field critical point while $\sigma$ does not, we expect that in the presence of noise the scale of typical fluctuations for $\varrho$ will be small compared to the scale of typical fluctuations for $\sigma$. Likewise, $\sigma$ will relax more slowly than $\varrho$ near the critical point, and will exhibit fluctuations on a longer length-scale, i.e. in the presence of fluctuations it will have a longer autocorrelation time than $\varrho$, and will be roughly spatially homogeneous over the length-scale on which $\varrho$ is correlated.  On the grounds of the latter statement, it is justified to solve \eref{eq:cant_eliminate_rho} at \emph{fixed} $\sigma$ and within a local-density approximation (i.e. we assume that $\varrho$ is relaxing in a locally homogeneous environment set by the slowly varying value of $\sigma$), in which case $\varrho$ simply undergoes a lattice version of Ornstein-Uhlenbeck relaxation to the mean value
\begin{align}
\bar\varrho=\varrho_0-\sqrt{3}K\nabla^2\sigma+\mathcal{O}(\sigma^2).
\end{align}
Straightforward analysis reveals that fluctuations of $\varrho$ around its mean value, $\vartheta(\tau)=\varrho(\tau)-\bar{\varrho}(t)$, obey
\begin{align}
\overline{\vartheta_j(\tau_1)\vartheta_k(\tau_2)}\sim \frac{\kappa}{\mathcal{N}}e^{-|j-k|/\ell_{\varrho}}e^{-|\tau_2-\tau_1|/\tau_{\varrho}},
\end{align}
where $\ell_{\varrho}\sim\sqrt{K}$ and $\tau_{\varrho}\sim1$ are the correlation length and correlation time of the massive field $\varrho$, respectively. Inserting the solution $\varrho=\bar\varrho+\vartheta$ into \eref{eq:need_to_simplify}, and keeping for now all terms that involve the fluctuations $\vartheta$ we obtain
\begin{align}
\label{eq:v_lang_fin}
\dot\sigma&=-\frac{h}{2}-r\sigma+K\nabla^2\sigma-g\sigma^3+\xi(\tau)\nonumber\\
&+2K\nabla^2\bar{\varrho}+2K\nabla^2\vartheta+\mathcal{O}(\bar{\varrho}^2)+\mathcal{O}(\bar{\varrho})\vartheta+\mathcal{O}(\vartheta^2).
\end{align}
The first and third terms on the second line, $2K\nabla^2\bar{\varrho}$ and $\mathcal{O}(\bar{\varrho}^2)$, contain terms that are either higher order in $\sigma$, $r$, or in gradients than other terms on the first line, and thus can be ignored.  The remaining terms depend on $\vartheta$; because the dynamics of $\sigma$ is slow and dominated by long wave-length fluctuations, we can approximate $\vartheta$ as spatially uncorrelated white noise with variance
\begin{align}
\overline{\vartheta_j(\tau_1)\vartheta_k(\tau_2)}\sim \frac{\kappa}{\mathcal{N}}\delta_{j,k}\delta(\tau_1-\tau_2).
\end{align}
In light of this approximation, the term $K\nabla^2\vartheta$ acts as an additional source of additive white noise---it can in principle be taken into account as a ($K$-dependent) renormalization of the noise $\xi(\tau)$ that is already present, but if we work to lowest-order in $K$ this renormalization can be ignored.  The term $\mathcal{O}(\bar\rho)\vartheta$ acts as a source of multiplicative noise; since $\bar\varrho$ is itself a small parameter, this noise can also be ignored in comparison to the already present white noise $\xi(\tau)$.  The term $\mathcal{O}(\vartheta^2)$ should be interpreted by writing $\vartheta(\tau)^2=f(\tau)+\chi(\tau)$, where $f$ is the average of $\vartheta^2$ and $\chi$ is its flucuations.  The average $f\sim\kappa/\mathcal{N}$ can be ignored for weak noise, while the fluctuations $\chi$ obey (using the fact that $\vartheta$ is a Gaussian variable)
\begin{align}
\overline{\chi_j(\tau_1)\chi_k(\tau_2)}\sim(\kappa/\mathcal{N})^2e^{-2|j-k|/\ell}e^{-2|\tau_1-\tau_2|/\tau_{\varrho}}.
\end{align}
As with $\vartheta$, $\chi$ can be interpreted as spatially-uncorrelated Gaussian white noise as far as the slow dynamics of $\sigma$ is concerned.  Since it has a variance that is parametrically smaller than that of $\xi$, it can once again be ignored.  With all of the terms on the second line of \eref{eq:v_lang_fin} dropped, we recover \eref{eq:vdot_noise} of the main text.

\

\section{Domain walls\label{sec:appendix_C}}

As described in the main text, the domain wall velocity can be determined by solving for the dynamics of a fictitous particle obeying the equation of motion
\begin{align}
\label{eq:particle_analogy}
K\ddot{\sigma}=-v\dot{\sigma}+\frac{\partial U(\sigma)}{\partial \sigma},
\end{align}
where
\begin{align}
U(\sigma)=\frac{1}{2}\big(r \sigma^2+\frac{1}{2}g \sigma^4+h \sigma\big).
\end{align}
Here the domain-wall velocity $v$ plays the role of a velocity dependent friction coefficient.  We seek solutions of \eref{eq:particle_analogy} for which the particle starts (with $\dot\sigma=0$) at the higher local maximum of $-U(\sigma)$ and comes to rest at the lower one. When $h$ is small the two local maxima have similar energies, and the friction coefficient $v$ must also be small for the particle to reach the top of the lower potential maximum.  Thus the trajectory is similar to that in the case of $h=0$, for which (from conservation of energy) $K\dot{\sigma}^2/2=U(\sigma)-U(\sigma_0)$, such that
\begin{align}
\label{eq:v}
\dot{\sigma}\approx\sqrt{\frac{2(U(\sigma)-U(\sigma_0))}{K}}.
\end{align}
The work done along this trajectory by the friction must be equal to the change in potential energy,
\begin{align}
\label{eq:c_of_e}
-v \int_{\sigma_-}^{\sigma_+} \dot{\sigma}\,d\sigma=U(\sigma_+)-U(\sigma_-)\approx h\sigma_0.
\end{align}
Inserting \eref{eq:v} into \eref{eq:c_of_e}, taking the integral, and using $U(\sigma_0)=r^2/4g$, we obtain the domain-wall velocity given in the main text,
\begin{align}
\label{eq:v_approx}
v\approx h\frac{3}{2}\sqrt{\frac{K g}{2r^2}}.
\end{align}


%


\end{document}